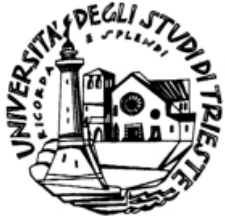

# Università degli Studi di Trieste

## Facoltà di Scienze Matematiche, Fisiche e Naturali

## Corso di Laurea in Fisica

### TESI DI LAUREA

---

# *The Cosmic Star Formation Rate: a theoretical approach*

---

*Laureando:*
Vincoletto Luca

*Relatrice:*
Ch. ma Prof.ssa
Matteucci Maria Francesca

Anno Accademico 2010-2011

# CONTENTS

Contents

## LIST OF FIGURES

List of Figures

## LIST OF TABLES

# INTRODUCTION

The rate at which galaxies have formed stars throughout the whole cosmic time is a fundamental clue to investigate the history of the assembling and evolution of structures in the Universe.

The *cosmic star formation rate (CSFR)*, defined like the comoving space density of the global star formation rate in a unitary volume of the Universe, is not a directly measurable quantity. It can be inferred only from the measurement of the luminosity density in different wavebands. The first determination of the CSFR has been made by Lilly et al. (1996), in three bands (2800 Å, 4400 Å and 1 $\mu m$) in the redshift range $0 < z < 1$. From that moment on, many studies using different indicators, have been conducted in the same wavelength range, for example Hogg et al. (1998) using the $[OII]$ line or Wilson et al. (2002), using the $UV$ luminosity density. These studies allowed to constrain the CSFR from $z = 0$ to $z = 1$, establishing that it is steadily increasing in that range.

Later on, and thanks to more and more deep surveys it has been possible to extend the determination of the cosmic star formation rate up to $z \sim 10$ (Madau et al. 1998, Hopkins 2004, Bouwens et al. 2008, Li 2008, Kistler et al. 2009, Bouwens et al. 2011, Ishida et al. 2011), even if the emerging picture seems less clear than at low redshift, and a clear trend of the CSFR is not yet well defined. In all of these works, the CSFR has been calculated on the basis of observational data.

On the other hand, also purely theoretical works are present in literature. For example Rowan-Robinson (1999) applied a parametric approach to analyze different histories of star formation to confront them with IR-submm data. Porciani and Madau (2001) have proposed several parameterizations some of which were taken from other authors (Madau and Pozzetti 2000 and Steidel et al. 1999), in order to be able to reproduce some observational quantities like number counts and redshift distribution of Gamma Ray Bursts (GRB). Moreover, Calura and Matteucci (2003) developed a model to compute the CSFR based on models of chemical and photometric evolution of galaxies, in the framework of the monolithic collapse scenario of galaxy formation, under the hypothesis of a *pure luminosity evolution (PLE)* of galaxies. Finally Menci et al. (2004) developed a semi-analytical model on the basis of the hierarchical clustering galaxy formation scenario.

Constraining the cosmic star formation rate at high redshift would be of paramount importance to put constraints is the galaxy formation scenario. In particular, to decide between the *Monolithic Collapse (MC)* and the *Hierarchical Clustering (HC)*, because the CSFR depends on both the star formation rate (SFR) in galaxies and on the evolution of the galaxy luminosity function.

In the MC scenario, spheroids and bulges formed at high redshift (e.g. $z > 2 - 3$) as the result of a violent burst of star formation, following the collapse of a gas cloud. After the collapse, galaxies evolve passively until present time (Larson 1974, Matteucci and Tornambé 1987). Moreover, Matteucci (1994) introduced in this scenario the assumption that more massive spheroids present higher star formation efficiencies than the less massive ones, and stop to form stars earlier. It is common to refer to this behavior as "downsizing". These assumptions allow us to reproduce the majority of chemical and photometric properties of local ellipticals.

On the other hand, in the HC scenario, spheroids and bulges formed as a series of subsequential mergers among gas-rich galaxies or with galaxies that have already stopped their star formation. This theory is based on gravitational instabilities in the distribution of the mass of the Universe, that is dominated by dark matter. Because of their lower Jeans mass, small structures, intended as dark matter halos, are the first to form. Then these small halos interact to form larger halos where larger galaxies are hosted. In this scenario galaxies form stars at lower rates than in the MC scenario, with more massive spheroids reaching their final mass at later times ($z \leq 1.5$) (White and Rees 1978, Baugh et al. 1998, Cole et al. 2000, Menci et al. 2004).

From the theoretical point of view, in order to compute the CSFR one needs to have the star formation histories (SFH) of galaxies of different morphological type on one hand, while on the other hand the evolution of the luminosity function of galaxies should be assumed. Obviously, these quantities have to be placed in the context of a cosmological model.

In this work of thesis, we have determined the CSFR from the theoretical point of view. We start in chapter 2 by describing detailed models of chemical evolution for galaxies of different morphological types (ellipticals, spirals and irregulars) from which we will derive the *star formation (SF)* histories. We adopt the models developed by Matteucci and Francois (1989) for spirals, Pipino and Matteucci (2004) for ellipticals and Bradamante, Matteucci, and D'Ercole (1998) for irregulars. Star formation histories obtained with chemical evolution models are then used to determine the photometric evolution of the galaxies, through a spectro-photometric code (GRASIL, Silva et al. 1998).

Chapter 3 is devoted to this subject. Here we describe the spectro-photometric code used and show the main results regarding spectra, luminosities and colors of ellipticals, spirals and irregulars.

The next step to compute the CSFR is the calculation of the luminosity density. For this purpose we need to know how galaxies are distributed in the Universe. This is possible thanks to the luminosity function (LF), that gives us the number of galaxies in a unitary volume of Universe in a luminosity bin. The luminosity function is usually represented through the functional form introduced by Schechter (1976). The LF is characterized by three parameters: the number density $\varphi^*$, the characteristic luminosity at the break $L^*$ and the slope $\alpha$. For luminosities fainter than $L^*$, it reduces to a power-law with slope $\alpha$; at brighter luminosities it gets an exponential shape. In this work we start with the local LF derived by Marzke et al. (1998), based on a local sample of galaxies, and determined in the B band for galaxies of different morphological type, and then we compute the evolution in luminosity of the galaxy at the break of the LF for each galactic type. Once we have the luminosity density we can calculate the CSFR by using: the star formation rate, the luminosity density and the mass-to-light ratio at each redshift.

In chapter 4 we calculate the CSFR under different evolutionary scenarios. First, we study the hypothesis of a pure luminosity evolution scenario. In this case, galaxies are supposed to form all at the same redshift and then evolve only in luminosity without any merging or interaction. In this case we follow the approach of Calura and Matteucci (2003).

Then we define scenarios in which the number density and the slope of the LF are assumed to vary with redshift.

All our results have been compared with data available in literature (Hopkins 2004, Li 2008, Kistler et al. 2009) and with other models. Then in chapter 5, as a consequence of he comparison between model results and data, some conclusions are drawn.

# THE CHEMICAL EVOLUTION OF GALAXIES

Chemical evolution models are the building blocks of this work. Starting from some observational constraints such as present day abundances it is possible to perform a detailed backward evolution of galaxies of different morphological type.

This allows us to give the time evolution of several quantities such as the star formation, the production rate of chemical elements and the chemical abundances in the stars and in the gas.

In this chapter, after a general description of the different morphological types of galaxies, we will give a detailed description of the models used together with some results coming from chemical evolution models.

## 2.1 THE MORPHOLOGICAL CLASSIFICATION

The study of galaxy evolution is a main topic of modern astrophysics. The capabilities offered from a wealth of observational facilities, allowed us to better understand the physical and observational properties of these objects. Starting from the first studies in the late seventies and early eighties, it has been possible to discover the Universe in his depth with increasing detail.

External galaxies were recognized as such only in the twenties of the last century. Before that they were supposed to be objects called *nebulae* and there was a so-called "great debate" between Shapley and Curtis on the fact that they were inside or outside our Galaxy. Hubble resolved the debate in 1922 in favor of Curtis, thanks to the discover of a Cepheid in the Andromeda galaxy (M31), that allowed him to measure correctly the distances.

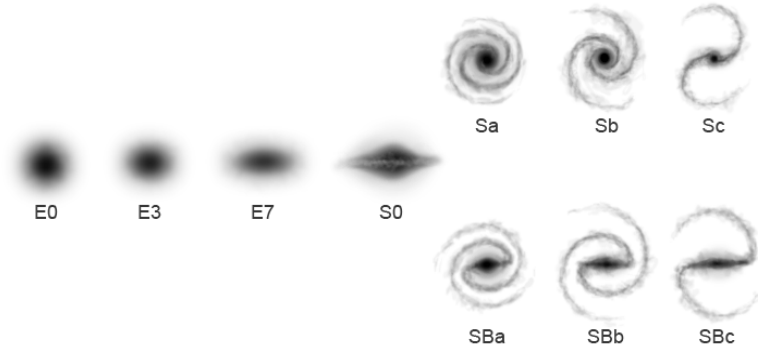

Figure 1: The Hubble "tuning fork" diagram

Although galaxies are very complex systems and their shapes and physical features are in general heterogeneous, several morphological classifications have been attempted. The most popular classification scheme is the one introduced by Edwin Hubble (1936). Although created a long time ago this classification is still widely used.

Beside defining the shape of galaxies, the Hubble sequence, shown in figure 1, has to be read also as a sequence of stellar population with the "earliest" types (from E0 to E7), i.e. the elliptical galaxies, characterized by very old, red stellar populations. The increasing prevalence of blue young stars, reaches its maximum at the opposite end of the fork, where late-type spirals and irregulars lie. However, nearly all galaxies contain very old stars, and the different proportions of red and blue stars arise from different star formation histories. From E0 to E7 we have the elliptical galaxies, which show no signs of hot young stars and have negligible gas or dust and therefore negligible star formation. They vary in shape from round to more elongated. The different label E$n$ describes the axial ratio $(b/a)$, according to the formula $n = 10 \cdot [1 - (b/a)]$, with $a$ and $b$ indicating major and minor axis respectively. An example is shown in fig. 2

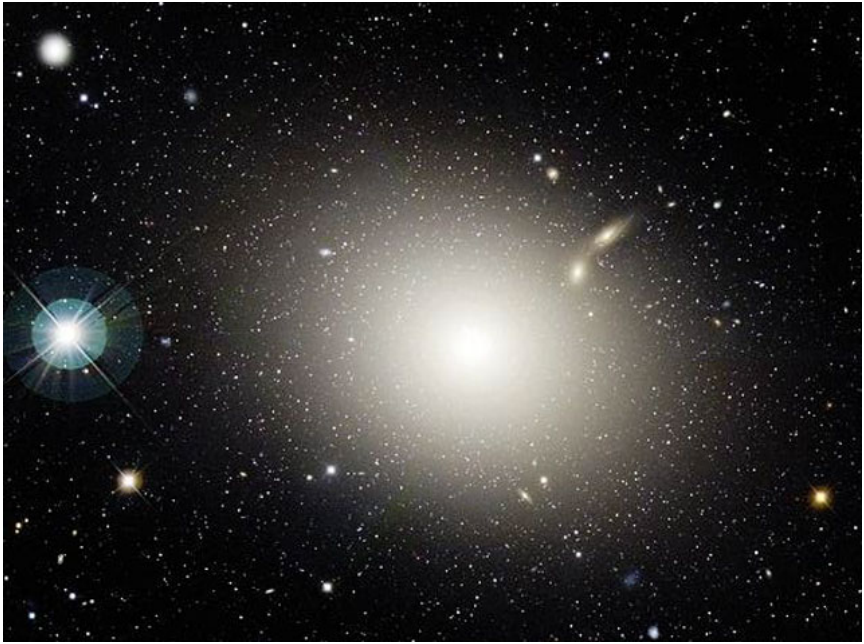

Figure 2: The elliptical galaxy M87. Optical image from Canada-France-Hawaii Telescope.

After the ellipticals, and before the junction with the spirals, in the middle of the Hubble fork, we find *lenticular galaxies* ($S0$). They are characterized by smooth central brightness, represented by the bulge (very similar to an elliptical galaxy), which is surrounded by a region of gradually declining brightness, in general characterized by a flat shape. In some cases their bulges are surrounded by a smooth structure whose shape is similar to a lens, from which this type of galaxies takes its name.

After lenticular galaxies the Hubble diagram bifurcates into two branches: the "normal" and the "barred" spiral galaxies. A normal spiral is characterized by a bright central structure resembling an elliptical galaxy, surrounded by a thin disk. From the center of spiral galaxies, central arms open up towards the outskirts: these structures have generally enhanced luminosity and are the regions of the disk with the strongest condensation of stars and gas. Barred spirals have instead a central bar-shape structure that surrounds their bulge. This division between normal and barred spirals can be further refined on the basis of three parameters:

1. The prominence of the central bulge and the ratio between the amount of light produced by the bulge and by the disk;

2. the tightness of the spiral arms;

3. the degree to which spiral arms are resolved into stars and individual emission nebulae.

Following this scheme normal spirals range from Sa to Sc and barred spirals from SBa to SBc, where Sa and SBa have the most prominent bulges and tight arms while Sc and SBc have small condensation of light in their bulges and arms are loosely wrapped. The Milky Way, (fig. 3) is classified as Sbc.

All the galaxies that can not be addressed to any of the described categories are classified as *Irregular*. Examples are the Large Magellanic Cloud (LMC) (fig. 4) and the Small Magellanic Cloud (SMC), that are classified as "dwarf" galaxies since their size is smaller than the Milky Way (with typical masses in the range $10^7 - 10^9$ M$_\odot$, Carroll 2006). Irregular galaxies are characterized by blue colors, sign of young, hot stars, although also elder stellar populations are recognizable. Irregulars, sometimes, may have very warped structures: in this case they are often referred as "peculiar". In many cases they are suffering interactions with other galaxies and this causes their distorted shape.

The classification of Hubble is not the only one although it is considered quite satisfactory. For example de Vaucouleurs (1959) found in the sequence particular structures like rings and spiral structures indicated as *r* and *s*, respectively. Another classification is the one of Morgan (1959) based on the stellar population that dominates bulge and disk respectively. In his classification galaxies are divided in *af, f, fg, gk* and *k* and the letters correspond to how much light from the bulge dominates the disk. Obviously, since galaxies have no sharp definition their spatial extent is usually defined through the *Holmberg's radius*, i.e. the radius of the isophote corresponding to a surface brightness of 26.3 $mag \cdot arcsec^{-2}$.

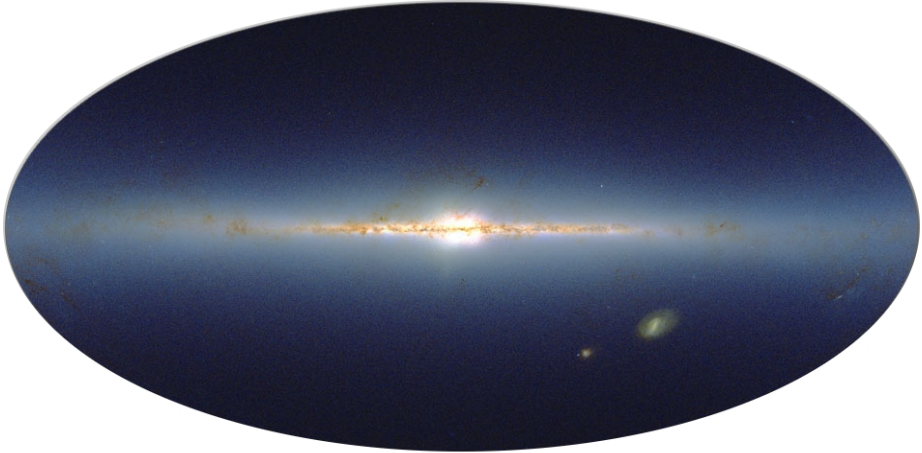

Figure 3: The Milky Way Galaxy. Infrared Atlas Image obtained as part of the Two Micron All Sky Survey (2MASS), a joint project of the University of Massachusetts and the Infrared Processing and Analysis Center/California Institute of Technology, funded by the National Aeronautics and Space Administration and the National Science Foundation.

Galaxies are often gathered in clusters but the majority of the systems in the visible Universe is isolated or composes small groups in the field, such as the Local Group (to which our Galaxy belongs). Ellipticals ($\sim 80\%$, Matteucci 2007)

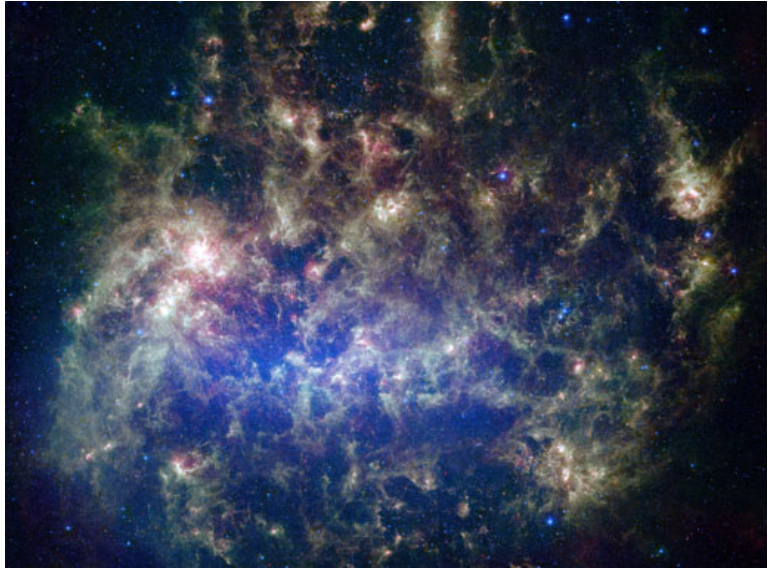

Figure 4: Infrared image of the Large Magellanic Cloud example of irregular galaxy. Image from Spitzer Space Telescope. NASA/JPL-Caltech/STScI

are mostly found in clusters while spirals and irregulars are usually found in the field.

## 2.2 CHEMICAL EVOLUTION OF GALAXIES

The baryonic matter in galaxies is continuously reprocessed by the stars and restored into the *interstellar medium (ISM)* through stellar winds and supernova (SN) explosions. The chemical evolution models allow us to calculate the time evolution of the production rate of the chemical elements and to follow the evolution in time and space of the abundances of these elements in the gas.

2.2.1   *Basic Principles*

The principal physical ingredients necessary to build up a model of chemical evolution can be summarized as:

- The initial conditions;

- the stellar yields;

- the stellar initial mass function (IMF);

- the star formation rate (SFR);

- the gas flows (infall, outflow, radial flows).

*The Initial Conditions*

They concern in particular the amount of gas present initially. The gas can be completely present at the time $t = 0$, or in other cases, it is progressively accumulated through an *infall law*. The rate of accretion is different for different objects and can be fast or slow depending on the type of galaxy we want to model. Another important condition regards the chemical composition of the gas that can be both pristine or pre-enriched. Usually, initial chemical abundances coming from the Standard Big Bang Nucleosynthesis are adopted, namely: $X_p = 0.76$, $Y_p = 0.24$ and $Z_p = 0$ respectively for the primordial mass abundances of H, $^4$He and metals, where it is useful to remember that in astrophysics the word "metal" refers to all the elements heavier than $^4$He.

*The Stellar Yields*

With the term stellar yields are usually indicated the masses of newly formed and preexisting elements produced and ejected by a star of mass $m$ (in units of solar mass) and metallicity $Z$. The yield per stellar generation (i.e. coeval stars) is instead defined as:

$$y_i = \frac{1}{(1-R)} \int_1^\infty m \, p_{im} \varphi(m) \, dm,$$ (2.1)

where $R$ represents the total mass, nuclearly processed and unprocessed which is ejected by a stellar generation and $p_{im}$ is the fraction of mass newly produced and expelled by a star of mass $m$:

$$R = \int_1^\infty (m - m_{rem}) \varphi(m) \, dm$$ (2.2)

where $m_{rem}$ is the mass of the stellar remnant (white dwarf, neutron star or black hole). $R$ is defined under the *Instantaneous Recycling Approximation* (IRA) (Tinsley 1980). This approximation states that all the stars more massive than $1 M_\odot$ die instantaneously while stars of $M \leq 1 M_\odot$ live forever.

Stellar yields can be separated in different contributions and specifically in yields from low and intermediate mass stars, massive stars and type Ia Supernovae.

*Low and Intermediate Mass Stars (LIMS)*

They are stars ranging from 0.8 to 8 $M_\odot$ and contribute through quiescent mass loss and the *Planetary Nebula (PN)* phase to the enrichment in $^4He$, $^{12}C$, $^{13}C$, $^{14}N$, $^{17}O$ and heavy $s$-process elements. The major sources of uncertainty affecting the nucleosynthesis of LIMS, are mainly related to the physical description of process occurring in stars, such as convection and mass loss during stellar evolution. Particularly uncertain are the predictions for the production of N, which is expected to be produced principally by LIMS during H-Burning in the CNO cycle. Its production depends on the quantity of C and O already present in the stellar envelope, which should have been produced by previous stars. N can also have a primary origin (i.e. formed starting from $H$ and $He$). In other words $N$ is primary if $C$ and $O$ have been produced *in situ*. This can happen during the Asymptotic Giant Branch phase of intermediate mass stars, when dredge-up occurs (Renzini and Voli 1981). The quantity of primary nitrogen is uncertain since it depends critically upon the treatment of convection and could also depend on rotation in massive stars as pointed out by Meynet and Maeder (2002). In this work the yields from LIMS are the ones calculated by van den Hoek and Groenewegen (1997) in which the initial metallicity of the stars is taken into account.

*Massive Stars*

With *massive stars* we indicate all the stars more massive than $8M_\odot$, that are responsible for the creation of the bulk of heavy element (with the exception of the Fe-peak ones), and in particular of oxygen which is the predominant element in the Universe after hydrogen and helium. Massive stars contribute to the ISM enrichment through type II SN explosions that originates from core collapse of single massive stars ($M > 8M_\odot$) leaving a neutron star or a black hole as remnant. Uncertainties in the yields from massive stars derive from the physical treatment of convection and from the uncertain nuclear reaction rates. The huge uncertainty concerns iron, since different studies (Thielemann et al. 1996, Woosley and Weaver 1995) give different values (differing by a factor of $\sim 3$) for stars with masses $> 15M_\odot$. On the other hand, for other elements like $\alpha$ elements ($^{16}O$, $^{24}Mg$) there is a substantial agreement among various authors. The yields for massive stars are taken from the work of Woosley and Weaver (1995).

*Type Ia Supernovae*

Type Ia SNe are commonly believed to originate from exploding white dwarfs in binary systems, although the nature of their progenitors is still matter of debate. The most popular model is the C-deflagration of a C-O white dwarf with a mass of $\sim$ 1.44 $M_{\odot}$ (Chandrasekhar mass) triggered by accretion of material from a companion, that can either be a red giant or another white dwarf. They are found in galaxies of all morphological type and, since they are present also in ellipticals, which have stopped forming stars several Gyrs ago, it is clear that their progenitors must be long-lived stars. Type Ia SNe are in charge for the production of iron ($\sim$ 0.7 $M_{\odot}$ each) along with traces of *Si and C*. SNe Ia are classified in this way due to their spectra, in which there is no sign of *H$\alpha$* absorption line, but *SiII* absorption line at $\lambda$ 6355 Å is present. In figure 5 it is shown a typical spectra of a Type Ia SN.

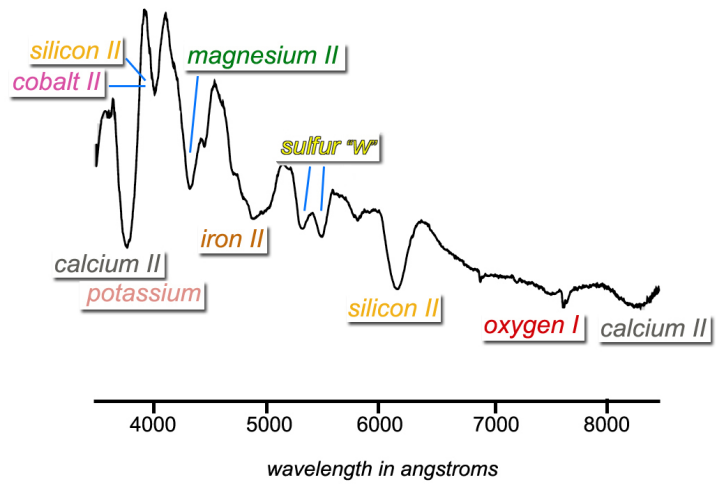

Figure 5: The Spectra of a Type Ia Supernova

There are two main scenarios which can lead to a Type Ia SN:

- THE SINGLE DEGENERATE SCENARIO

  In this scenario (Nomoto et al. 1984) there is the C-deflagration in a C-O white dwarf when reaching the Chandrasekhar mass after accreting material from a red giant companion. The progenitors of C-O WDs lie in the range $0.8 - 8\ M_\odot$. The minimum timescale for the appearance of the first Type Ia SNe is the lifetime of the most massive secondary star (0.03 Gyr for an $8\ M_\odot$).

- THE DOUBLE DEGENERATE SCENARIO

  It consists in the merging of two C-O WDs, due to loss of angular momentum occurring as a consequence of gravitational wave radiation, which then explode by C-deflagration when the $M_{Ch}$ is reached (Iben and Tutukov 1984). In this case the minimum timescale for the appearance of Type Ia SNe is the lifetime of the most massive secondary star, plus the gravitational time delay which depends on the original separation of the two WDs and which is computed according to Landau and Lifshitz (1962). It is useful to remind that the gravitational time delay can be as long as several Hubble times and as short as 1 Myr (Greggio 2005).

In this work, for SN Ia we consider the yields from Iwamoto et al. (1999) and in particular their W7 model.

*The Stellar Initial Mass Function (IMF)*

The IMF, i.e. the number of stars born in the mass interval [*m, m+dm*], is usually expressed as:

$$\varphi(m)dm = \varphi_0 m^{-(1+x)}dm, \tag{2.3}$$

where $\varphi_0$ is a normalization constant and $x$ is the slope. The determination of the IMF can be obtained starting from *n(m)*, namely the distribution of main sequence stars obtained empirically counting the number of stars per unit of magnitude in the solar neighborhood (a 1 *Kpc* cylinder centered on the Sun). These counts must be corrected for the presence of post-main sequence stars, and then the mean number of stars per $pc^{-2}$ is calculated considering stars born at every height relative to the galactic plane. This is necessary to pass from absolute magnitude to mass and it requires the knowledge of a mass-luminosity relation. Unfortunately such relation is not uniquely determined since stars evolve while they are in main sequence and therefore for a given chemical composition, it is possible to have different relations. Once $n(m)$ is determined then $\varphi(m)$ can be calculated. Assuming that the IMF is constant in time we can write

- In the range $0.1 < M/M_\odot \leq 1$:

$$n(m) = \varphi(m)\langle\psi\rangle t_{Hubble},\tag{2.4}$$

  where $\langle\psi\rangle$ is the mean value in the past of the star formation rate and $t_{Hubble}$ is the galactic lifetime (13-14 Gyr).

- In the range $M > 2M_\odot$:

$$n(m) = \varphi(m)\psi(t_{Hubble})\tau_m,\tag{2.5}$$

  where $\psi(t_{Hubble})$ is the current star formation rate and $\tau_m$ is the lifetime of a star of a given mass.

- In the range $1 < M/M_\odot < 2$ it is not possible to apply the rules described before, but it is possible to define a parameter $b$ given by:

$$b = \frac{\psi(t_{Hubble})}{\langle\psi\rangle},\tag{2.6}$$

which, in our Galaxy ranges in the interval $[0.5 - 1.5]$, namely the SFR is changed by no more than a factor of $\sim 2$ from the past till now. This is a very important constraint on the SFR in the solar vicinity.

The widest used IMF is the one of Salpeter (1955) that is characterized by a single-slope power law with $x = 1.35$. Other common IMFs found in literature are the one of Scalo (1986) which is a double-slope IMF with $x = 1.35$ for $0.1 \leq M/M_\odot \leq 2$ and $x = 1.7$ for $2 < M/M_\odot \leq 100$, and more recently the Kroupa (2001) IMF and the Chabrier (2003) IMF, for the Milky Way. The Scalo IMF has been demonstrated to be the best in reproducing the observational constraints in the solar neighborhood, whereas the Salpeter has to be preferred for nearby galaxies (Lanfranchi and Matteucci 2003). In this work, it has been used the IMF of Salpeter (1955).

*The Star Formation Rate (SFR)*

The star formation rate is one of the most important drivers of galactic chemical evolution since it describes the rate at which the gas is turned into stars in galaxies. Since the physics of the star formation process is still not well known, several parametrization are used to describe the SFR. A common aspect to the different formulations of the SFR is that they include a dependence upon the gas density. The most famous and most widely adopted is the Schmidt (1959) law:

$$SFR = \nu \sigma_{gas}^k,$$ 
(2.7)

Schmidt suggested a value of 2 for the parameter $k$, but Kennicutt (1998) suggested that the best fit to the observational data on spiral disks and starburst galaxies is obtained with an exponent $k = 1.4 \pm 0.15$, as shown in fig. 6.

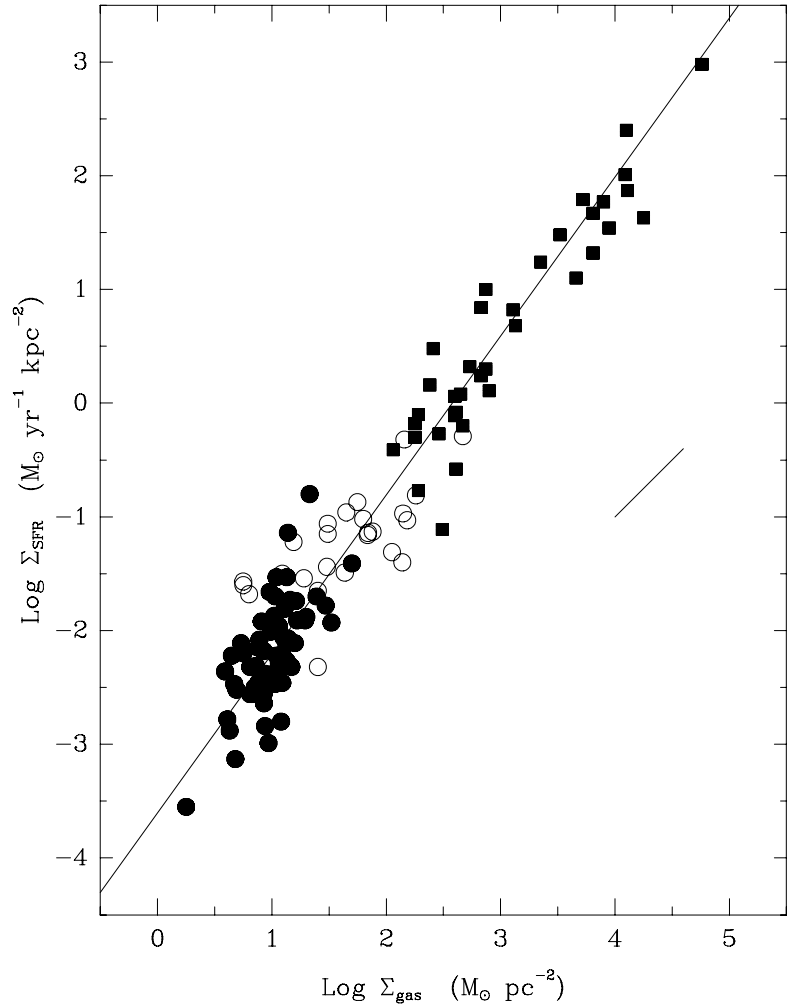

Figure 6: Observational correlation between the gas surface density and the SFR for normal
disks (circles) and starburst galaxies (squares). The line is a least square fit with $k$=1.4.
(Kennicutt 1998).

An exponentially decreasing SFR provides an easy to handle formula:

$$SFR = \nu e^{-t/\tau_*},\tag{2.8}$$

with $\tau_* = 5 - 15$ Gyr in order to obtain a good fit to the properties of the solar neighborhood (Tosi 1988) and $\nu = 1 - 2$ Gyr$^{-1}$ being the so-called efficiency of star formation, expressed as the inverse of the timescale of star formation.

*Gas Flows*

Gas flows include phenomena such as infall, outflow and radial flows of gas. The easiest way to describe the infall rate *WR* is through an exponential law:

$$IR \propto e^{-t/\tau},\tag{2.9}$$

where $\tau$ indicates the timescale of the infall and is a free parameter. For what concerns the outflow, or the galactic wind it is considered proportional to the SFR according to:

$$WR = -\lambda\psi,\tag{2.10}$$

where *WR* is the rate at which mass is lost in the galaxy, and the wind is assumed to set up when the thermal energy of the gas exceeds its binding energy. Also in this case $\lambda$ is a free parameter, tuned (like $\tau$) to reproduce the major observational constraints.

### 2.2.2 *Basic Equations*

In this section we will describe the models used in this work, starting from their common features and then going into detail for what concerns their peculiarity. In all the models, the solution of the equations is numerical. This approach allows to relax the instantaneous recycling approximation. If $\sigma_g(t)$ is the surface gas mass density in a given galaxy at time $t$, $\sigma_{tot}(t_g)$ is the total surface mass density of the galaxy at the present time $t_g$ and $X_i(t)$ is the abundance by mass of an element $i$ at the time $t$, then we can define:

$$G_i(t) = \frac{\sigma_g(t)}{\sigma_{tot}(t_g)} \cdot X_i(t), \qquad (2.11)$$

as the mass fraction of gas in the form of an element $i$ at the time $t$. The most general chemical evolution equation can be written in the following way:

$$
\begin{aligned}
\dot{G}_i(t) = & -\psi(t) X_i(t) \\
& + \int_{M_L}^{M_{Bm}} \psi(t - \tau_m) Q_{mi}(t - \tau_m) \varphi(m) \, dm \\
& + A \int_{M_{Bm}}^{M_{BM}} \varphi(m) \\
& \cdot \left[ \int_{\mu_{min}}^{0.5} f(\mu) \psi(t - \tau_{m2}) Q_{mi}(t - \tau_{m2}) \varphi(m) \, d\mu \right] dm \\
& + (1 - A) \int_{M_{Bm}}^{M_{BM}} \psi(t - \tau_m) Q_{mi}(t - \tau_m) \varphi(m) \, dm \\
& + \int_{M_{BM}}^{M_U} \psi(t - \tau_m) Q_{mi}(t - \tau_m) \varphi(m) \, dm \\
& + X_{Ai} A(t) - X_i W(t).
\end{aligned}
\qquad (2.12)
$$

All the variables in the equation above are normalized to $\sigma_{tot}(t_g)$. The first term on the right hand side, including the star formation rate $\psi(t)$, represents the rate at which each element disappears from the ISM owing to the star formation. The second term is the rate at which each element is restored to the ISM by single stars with masses in the range $[M_L, M_{Bm}]$, where $M_L$ is the min-imum mass contributing, at a given time $t$, to chemical enrichment ($\simeq 0.8 M_{\odot}$) and $M_{Bm}$ is the minimum binary mass allowed for binary systems to give rise to Type Ia SNe (3 $M_{\odot}$, Matteucci and Greggio 1986). The quantities $Q_{mi}(t - \tau_m)$ indicates the fraction of mass restored by the stars in form of an element $i$ that can be produced or destroyed in stars or both. This is the so-called *"production matrix"* (Talbot and Arnett 1973). The third term represents the enrichment due to binaries which become Type Ia SNe, i.e. all the binary systems with total mass in the range between $M_{Bm}$ and $M_{BM}$, with the first term defined as before and the second that can be at maximum equal to 16 $M_{\odot}$. The model adopted for the progenitors of the Type Ia SNe is the single degenerate (see pag. 24). The parameter $A$ represents the fraction of binary stars giving rise to Type Ia SNe, and although its real value is unknown it is fixed in order to reproduce the observed present time SN Ia rate. This parameter obviously depends on the adopted IMF and, in general, values between 0.05 and 0.09 are adopted. This ensures to be able to reproduce the actual SN Ia rate both in the Milky Way and in other galaxies. It is worth noting that, in this term, the quantities $\psi$ and $Q_{mi}$ refer to the time $(t - \tau_{m2})$, where $\tau_{m2}$ indicates the lifetime of the secondary star of the binary system, which regulates the explosion timescale of the system. With this approach, only binaries giving rise to a Type Ia SN are treated separately, whereas stars in other binary systems are treated as single stars. The coefficient $\mu = M_2/M_B$, is the ratio between the mass of the secondary component and the total mass of the binary system, while $f(\mu)$ is the distribution function of this ratio. It is calculated on a statistical basis indicating that values close to 0.5 are preferred. Its analytical expression can be written as:

$$f(\mu) = 2^{(1+\gamma)}(1+\gamma)\mu^{\gamma}, \qquad (2.13)$$

with $\gamma = 2$ as a parameter (Greggio and Renzini 1983) and $\mu_{\min}$ is the minimum mass fraction contributing to the SNIa rate at the time $t$, and is given by:

$$\mu_{min} = max \left[ \frac{M_2(t)}{M_B}, \frac{M_2 - 0.5M_B}{M_B} \right].$$ (2.14)

The fourth term represents the enrichment due to stars in the range $[M_{Bm}, M_{BM}]$ which are single or, if binaries, do not produce a SNIa event. All the stars in this range with mass greater then $8M_\odot$ are assumed to explode as core collapse of single massive stars. The fifth term represents the contribution to the chemical enrichment coming from stars more massive than $M_{BM}$ (supposed to explode like type II and $I_{b/c}$ SNe). In all the models the upper mass limit contributing to chemical enrichment is assumed to be 100 $M_\odot$. These are the first contributors to the chemical enrichment, especially at low metallicities, since their lifetime is very short and are the first stars to die. Finally the last two terms are the rate of accretion of matter with primordial abundances $X_A$, and the outflow rate for the element $i$, respectively.

The function $\tau_m(m)$ used to describe the stellar lifetimes is the one introduced by Padovani and Matteucci (1993):

$$\tau_m \left[ Gyr \right] = \begin{cases} 10^{\left[ 1.338 - \sqrt{1.79 - 0.2232 \cdot (7.764 - logM)} \right] \big/ 0.1116} - 9 & \text{if } M \leq 6.6 \ M_\odot \\ 1.2 \cdot M^{-1.85} + 0.003 & \text{if } M > 6.6 \ M_\odot \end{cases}$$ (2.15)

## 2.3 MODELS ADOPTED IN THIS THESIS

In the following we describe the specific chemical evolution models and histories of star formation adopted in this thesis.

2.3.1 *Spiral Galaxies*

The chemical evolution of spiral galaxies is studied by using the one-infall model of Matteucci and Francois (1989), developed to reproduce the observational constraints of the Milky Way. This model belongs to the class of the so-called *"infall model"*, namely the disk is thought as gradually forming by infall of primordial gas. In this case the halo is treated as an homogeneous sphere of primordial gas, evolving as a one-zone model. Instantaneous mixing of gas is assumed, but the instantaneous recycling approximation is relaxed. The galactic disk is supposed to have formed on timescales increasing with the galactocentric distance, being larger at larger radii, in agreement with the dissipative models of galaxy formation of Larson (1976). The gas accumulates faster in the inner than in the outer region, according to the so-called *"inside-out"* scenario. The process of disk formation is much longer than the halo and the bulge formation (they should form on timescales no longer than 1 Gyr, Matteucci (2001)) with typical timescales varying from $\sim$ 2 Gyr in the inner disk, $\sim$ 7 Gyr in the solar region and up to $15 - 20$ Gyr in the outer disk. This mechanism is important to reproduce the observed abundance gradients. The spiral disk is approximated by several independent rings, 2 kpc wide, without exchange of matter between them. In this work the disk has been set to extend from 2 to 16 *Kpc*.

The basic equation for the spiral model is eq. 2.12, without considering, in this case, the outflow term $W(t)$. The infall rate is defined as:

$$A_i(t) = \frac{dG_i(r,t)}{dt} = \frac{A(r) \cdot (X_A)_i \cdot e^{-t/\tau}}{\sigma_{tot}(r,t_G)}. \tag{2.16}$$

The quantity $A(r)$ is derived in order to reproduce the current total mass surface density distribution in the halo and along the disk and is defined by:

$$A(r) = \frac{\sigma_{(r,t_G)}}{\tau(r)[1 - e^{-t_G/\tau(r)}]}, \tag{2.17}$$

| $M_D(M_\odot)$ | $\tau_{inf}\ (Gyr)$ | $\nu\ (Gyr^{-1})$ |
|:---:|:---:|:---:|
| $4 \times 10^{10}$ | 7.0 | 1.0 |

Table 1: Model parameters for a Milky Way-like spiral galaxy as defined in this work. $M_D$ is the baryonic mass of the disk, $\tau_{inf}$ is the infall timescale for the solar neighborhood, $\nu$ is the star formation efficiency.

and depends on the assumed current total surface mass density in the disk. For the star formation rate the formulation adopted is the following:

$$\psi(r,t) = \nu \cdot \left[ \frac{\sigma(r,t)}{\widetilde{\sigma}(r_\odot,t)} \right]^{2(k-1)} \cdot \left[ \frac{\sigma(r,t_G)}{\sigma(r,t)} \right]^{(k-1)} \cdot G^k(r,t), \qquad (2.18)$$

where $\nu$ is the SF efficiency, set in this case to 1 $Gyr^{-1}$, $\sigma(r,t)$ is the surface mass density at a given radius and at a given time, $\widetilde{\sigma}(r_\odot,t)$ is the total surface mass density at a particular distance from the Galactic center (i.e. at the position of the Sun equal to 8.5 $Kpc$). Finally $\sigma(r,t_G)$ is the surface mass density at present time and $G(r,t)$ is the normalized surface density at time t. The exponent $k$ is set equal to 1.4 to reproduce the observational data. This formulation of the SFR is similar to that of Matteucci and Francois (1989). In this work we do not adopt a threshold density for the star formation in the disk as suggested in other works (Chiappini, Matteucci, and Gratton (1997) or Kennicutt 1998) The main parameters regarding the baryonic mass of the galaxy, the infall timescale (i.e. the time in which half of the mass of the system is accumulated in the galaxy) and the star formation efficiency (i.e. the rate at which gas is converted into stars) are summarized in table 1.

2.3.2 *Elliptical Galaxies*

The model used in this work is the model originally developed by Matteucci and Tornambé (1987), and subsequently modified by Martinelli, Matteucci, and Colafrancesco (1998) and Pipino and Matteucci (2004). This model is developed in the spirit of the Monolithic Scenario in which spheroids are formed at high $z$ as a result of a sudden collapse of a gas cloud of pristine chemical composition. The gas is supposed not to be already assembled at $t = 0$, but it is supposed to assemble in a finite timescale. Star formation is supposed to begin from $t = 0$, i.e. while the infall is occurring. The galaxy can be modeled both as a one-zone object and as a multi zone object in which the galaxy is divided into several concentric shells independent of each-other. In the first case the whole galaxy shares the same characteristics, while in the second one each concentric shell can have different characteristics (such as infall timescale, SF, radius, etc.). The main equation of the model is identical to eq. 2.12. In this picture the galaxy is supposed to suffer a strong initial burst of star formation, that is supposed to stop as soon as the galactic wind develops. The galactic wind is assumed to start when the energy injected in the ISM by SNe and stellar winds exceeds the potential energy of the gas. In this sense the evolution of elliptical galaxies crucially depends on the time at which the galactic wind occurs, namely $t_{GW}$. From that moment on the galaxy is supposed to evolve passively.

The condition for the onset of a wind can be written as:

$$(E_{th})_{ISM} \geq E_{B_{gas}},$$ (2.19)

where $(E_{th})_{ISM}$ and $E_{B_{gas}}$ are the thermal and binding energy of the interstellar gas. The thermal energy of gas due to SN and stellar wind heating is:

$$(E_{th})_{ISM} = E_{th_{SN}} + E_{th_w},$$ (2.20)

with:

$$E_{th_{SN}} = \int_0^t \epsilon R_{SN}(t')\, dt', \qquad (2.21)$$

and:

$$E_{th_w} = \int_0^t \int_{12}^{100} \varphi(m)\psi(t')\epsilon_w\, dm\, dt', \qquad (2.22)$$

for the contribution from SN and stellar winds respectively. The quantity $R_{SN}$ represents the SN rate (both II and Ia). The quantities $\epsilon_{SN} = \eta_{SN}\epsilon_0$ with $\epsilon_0 = 10^{51}\ erg$, and $\epsilon_w = \eta_w E_w$ are the efficiencies for the energy transfer from SN II and Ia into the ISM. The coefficient $\epsilon_0 = 10^{51}\ erg$ is the typical SN blast wave energy, and $E_w = 10^{49}\ erg$ is the typical energy injected into the ISM by a 20 $M_\odot$ star, taken as representative. The parameters $\eta_w$ and $\eta_{SN}$ represents respectively the fraction of the energy blast and of the typical energy injected into the ISM. They can be assumed as free parameters or be calculated from the result of the evolution of a SN remnant in the ISM and the evolution of stellar winds (Martinelli, Matteucci, and Colafrancesco (1998); Pipino, Matteucci, et al. (2002)). The total mass of the galaxy is expressed as:

$$M_{tot}(t) = M_*(t) + M_{gas}(t) + M_{dark}(t), \qquad (2.23)$$

where the three terms represents the stellar, the gaseous and the dark matter mass, respectively. The dark matter component is assumed to be present in a diffuse halo ten times more massive than the baryonic component of the galaxy and is supposed to be distributed within a scale radius equal to $R_{dark} = 10\ R_{eff}$ (Matteucci 1992).

The total baryonic mass can be defined as:

$$M_L = M_* + M_{gas}(t). \qquad (2.24)$$

The binding energy of the gas can be described, following Bertin et al. (1992) as:

$$E_{B_{gas}}(t) \ = \ W_L(t) + W_{LD}(t),$$ (2.25)

where:

$$W_L(t) \ = \ -0.5 \cdot G \cdot \frac{M_{gas}(t) M_L(t)}{r_L},$$ (2.26)

represents the potential well due to the luminous matter, whereas:

$$W_{LD} \ = \ -\frac{GS}{2\pi} \cdot (1 + 1.37S) \cdot \frac{M_{gas}(t) M_{dark}}{r_L},$$ (2.27)

is the potential well due to the interaction between dark and luminous matter, and $S \ = \ r_L/r_D$ is the ratio between the effective radius and the radius of the dark matter core.

The SFR is assumed to be:

$$\psi \ [M_\odot \cdot yr^{-1}] \ = \ \nu M_{gas},$$ (2.28)

where $\nu$ is the star formation efficiency, in the simple hypothesis that the efficiency of star formation goes like $M_L^{-\gamma}$ ($\gamma = -0.11$, Arimoto and Yoshii 1987), owing to the fact that the SF efficiency is just the inverse of the timescale for star formation:

$$\nu \ = \ \tau_{SF}^{-1},$$ (2.29)

and that:

$$\tau_{SF} \propto min(\tau_{coll}, \ \tau_{ff}),$$ (2.30)

with the collision timescale defined as:

$$\tau_{coll} = \lambda_J / v, \tag{2.31}$$

where $\lambda_J$ is the Jeans length and $v$ is the velocity dispersion. And the free-fall timescale defined as:

$$\tau_{ff} = (3\pi / 32 G\rho)^{1/2}, \tag{2.32}$$

where $G$ is the gravitational constant and $\rho$ is the density of the gas mass. Since dynamical timescales are longer for more massive galaxies, the efficiency of SF should decrease with galactic mass. The efficiency $v$, coupled with the increase of the potential well as $M_L$ increases leads to the fact that more massive galaxies form stars for a longer period before suffering a galactic wind. This description is commonly referred as the *"classic wind"* scenario. This has been invoked to explain the observed mass-metallicity relation, but to the contrary, to explain the $[Mg/Fe] - \sigma$ relation inferred for the nuclei of bright ellipticals, an efficiency of star formation increasing with luminous mass seems to be required (Matteucci 1994). This is also in agreement with the so-called *downsizing*, where the most massive objects are supposed to form stars in a shorter timescale and with an increasing efficiency. In this work, since we are interested in reproducing the characteristics of a typical elliptical galaxy, we consider only one star formation efficiency as stated in table 2, where $M_{tot}$ is the total baryonic mass, $R_{eff}$ is the effective radius of the galaxy (i.e. the radius at which one half of the total light of the galaxy is emitted interior to this radius), $\tau_{inf}$ is the infall timescale, $v$ is the star formation efficiency.

| $M_{tot}(M_\odot)$ | $R_{eff}$ $(Kpc)$ | $\tau_{inf}$ $(Gyr)$ | $\nu$ $(Gyr^{-1})$ |
|---|---|---|---|
| $10^{11}$ | 3.0 | 0.3 | 15.0 |

Table 2: Model parameters for a typical elliptical galaxy as defined in this work. $M_{tot}$ is the baryonic mass, $R_{eff}$ is the effective radius, $\tau_{inf}$ is the infall timescale and $\nu$ is the star formation efficiency.

### 2.3.3 *Irregular Galaxies*

Irregular galaxies (i.e. dwarf irregulars and blue compact) are very important objects to study galaxy evolution since they are easier to model because of their generally small sizes and simple structure. Their stellar populations appear to be mostly young, their metallicity is low and their gas content is large. All these features indicate that these galaxies are poorly evolved objects, and may have undergone discontinuous, gasping or continuous, but lower than other morphological types, star formation activity. Also in this case the fundamental equation for their study is eq. 2.12 and both terms of infall and outflow are included. In fact, irregular galaxies are assumed to assemble all their gas by means of a continuous infall of pristine gas. Furthermore, these galaxies, are allowed to develop a galactic wind (Bradamante, Matteucci, and D'Ercole 1998). The accretion term is given by:

$$A_i = C \frac{(X_i)_{inf} \cdot e^{-t/\tau}}{M_L} \tag{2.33}$$

where $(X_i)_{inf}$ is the abundance of the element $i$ in the infalling gas, assumed to be primordial; $\tau$ is the timescale of mass accretion (set in this work equal to 1 Gyr) and $C$ is a constant obtained by imposing to reproduce $M_L$ at the present time $t_G$. The star formation in the model, can proceed both in bursts separated by a more, or less, long quiescent period (Blue Compact Galaxies, *BCG*) or at a low regime but continuously (Magellanic irregulars). It is also possible, of course, to have a regime in which the duration of the bursts is longer than their separation (*gasping* SF). In this work it has been chosen a continuous and constant star formation from the beginning till now with a continuous burst. The SFR can be expressed by:

$$\psi(t) = \nu M_{gas}(t), \tag{2.34}$$

where, as said before, $\nu$ is the star formation efficiency expressed in units of Gyr$^{-1}$ and $M_{gas}(t)$ represents the gas mass at the time $t$. Also in irregular galaxies it is possible to develop a galactic wind which occurs, as for elliptical galaxies, when the thermal energy of the gas overcomes its binding energy. The rate of gas lost via galactic wind for the $i$ element is:

$$W_i(t) = X_i(t)W(t) = X_i(t)w_i\psi(t), \tag{2.35}$$

where $w_i$ is a free parameter which describes the efficiency of the galactic wind. The value of $w_i$ has been in general assumed to be the same for different elements, although some authors (Yin, Matteucci, and Vladilo 2011) claimed that a *differential* wind i.e. different $w_i$ for different elements, best reproduces the properties of irregulars. In this work we made such a choice setting $w_i$ equal to 0.3 for H, D, $^3$He and $^4$He, and equal to 1.0 for the remaining chemical species. Table 3 summarizes the adopted parameters for the irregular galaxy model used in this work, regarding the baryonic mass $M_{tot}$, the infall timescale $\tau_{inf}$ and the star formation efficiency $\nu$.

| $M_{tot}(M_\odot)$ | $\tau_{inf}$ $(Gyr)$ | $\nu$ $(Gyr^{-1})$ |
|---|---|---|
| $10^{10}$ | 0.5 | 0.1 |

Table 3: Model parameters for a typical irregular galaxy as defined in this work. $M_{tot}$ is the baryonic mass, $\tau_{inf}$ is the infall timescale, $\nu$ is the star formation efficiency.

## 2.4 RESULTS FROM CHEMICAL EVOLUTION MODELS

In this thesis we are mostly interested in the SFRs of galaxies in order to model the cosmic star formation rate. To do that we have to adopt galaxy evolution models that reproduce the majority of observed constraints of local galaxies; in particular the chemical abundances. In what follows we present some results concerning SFR, metallicity and chemical abundances for the different morphological types of galaxies studied.

### 2.4.1 *The Star Formation Rate*

In figure 7 the star formation histories adopted in this work are shown. As one can see a typical elliptical experiences a high burst of star formation lasting for about 0.6 *Gyr* with a maximum peak of more than $\sim 100 \; M_\odot yr^{-1}$. After such a period, the star formation ceases abruptly owing to the onset of the galactic wind. The spiral is characterized by a continuous SFR with a large peak around $1 \sim 2 \; Gyr$ of less than $\sim 10 \; M_\odot yr^{-1}$ and a present time value of $\sim 2 \; M_\odot yr^{-1}$ (for the Milky Way). Finally, the irregular galaxy forms stars at a rate smaller than the previous morphological types, with an increasing SFR that reaches a maximum of less then $1 \; M_\odot yr^{-1}$, with an even smaller present value, due to the onset of the galactic wind.

### 2.4.2 *Abundances and Abundances Ratios*

Here some results concerning the chemical abundances and the abundances ratios for all the morphological types are presented. These abundances are derived from the solution of the integro-differential equations presented in section 2.2.2.

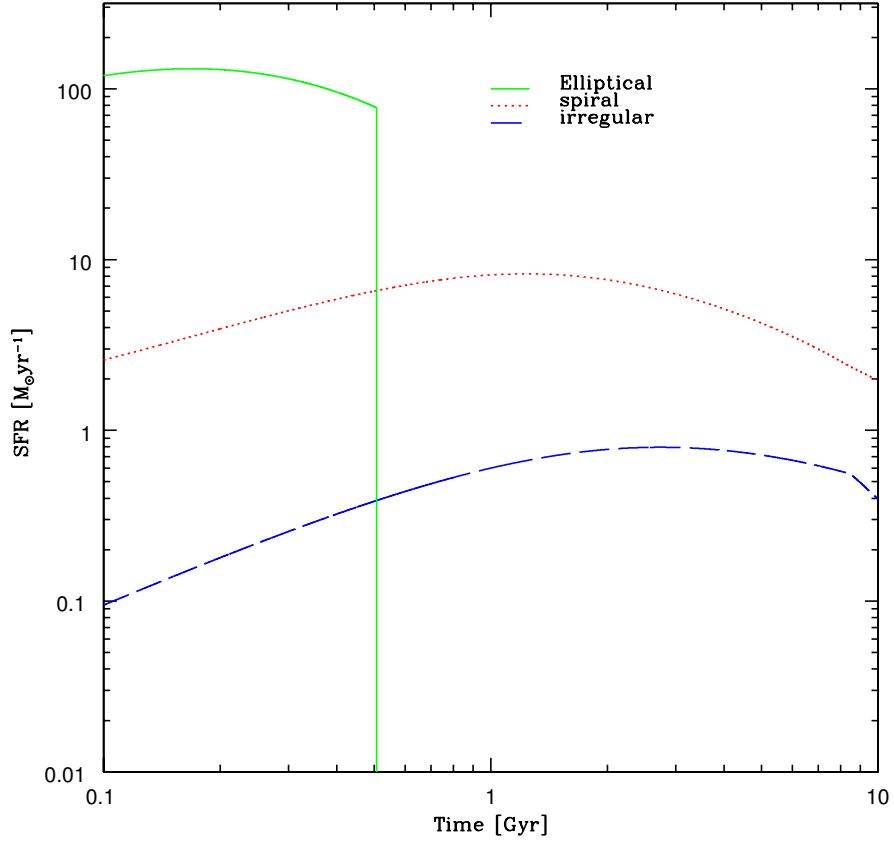

Figure 7: Star formation rates as a function of time, for a typical elliptical, spiral and irregular galaxy.

In figure 8 it is possible to see the time evolution of the abundance of oxygen ($[O/H]$)[1] for different galaxies. It is clear that the morphological type that exhibits the higher abundance are ellipticals. They experience a strong increase of oxygen abundance until $\sim 0.6$ Gyr when the onset of the galactic wind stops it from growing. For spirals the increase is continuous until present time. Irregulars show a lower abundance of oxygen than the other morphological types and is clear the decrease of the abundance of this element when the galactic wind sets on.

---

[1] For any chemical element $X$ we define: $[X/H] = log(X/H)_* - log(X/H)_\odot$ where $log(X/H)_*$ is the abundance ratio of an element relative to hydrogen, in the object we are studying and $(X/H)_\odot$ is the same abundance ratio in the Sun.

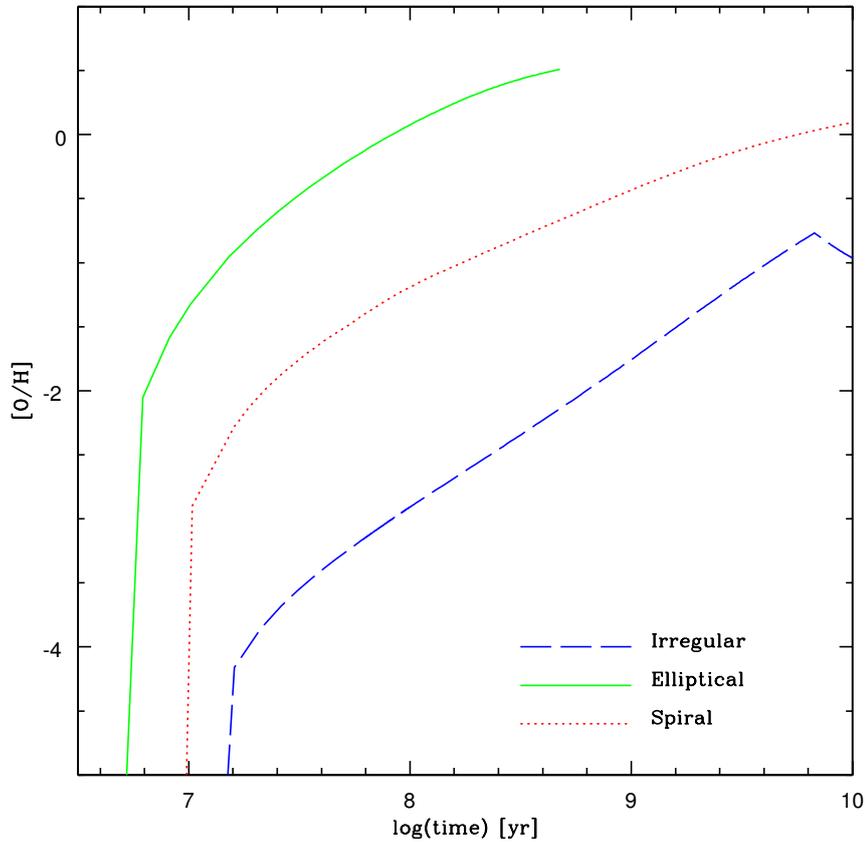

Figure 8: [O/H] vs Time for a spiral, an elliptical and an irregular galaxy.

The same trend is shown in figure 9 for the time evolution of the abundance of $[Fe/H]$, with again ellipticals reaching their peak at very early times (due to their strong SF at early phases) while spirals and irregulars show a slow but continuous increase of this abundance. The final decrease for irregulars is due again to the galactic wind.

Of particular importance in chemical evolution studies is the analysis of the abundance ratios. In fact, while absolute abundances usually depend on all the model assumptions, abundance ratios depend mainly on nucleosynthesis, stellar lifetimes and IMF. Abundances ratios, can therefore be used as cosmic clocks if they involve two elements formed on quite different timescales.

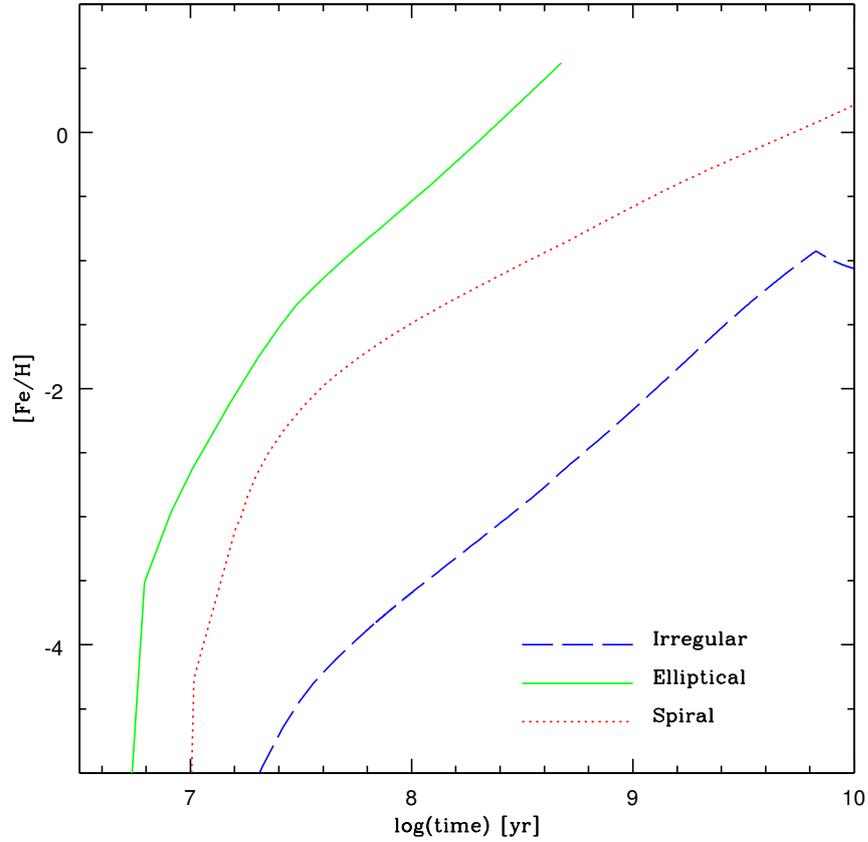

Figure 9: [Fe/H] vs Time for a spiral, an elliptical and an irregular galaxy.

In figure 10 we show the trend of the $[O/Fe]$ $vs$ $[Fe/H]$ ratio, with oxygen taken as representative of the $\alpha$ elements. With the term $\alpha$ elements we indicate a set of chemical elements, like for example $^{16}O$, $^{20}Ne$, $^{24}Mg$ and $^{28}Si$ produced by means of a series of chemical reactions involving $\alpha$ particles that occurs in stellar interior. It is possible to note that ellipticals show a trend with supersolar values for a wide range of $[Fe/H]$ values, and then a decline. The decline has to be ascribed to the different histories of star formation experienced by ellipticals and spirals. In fact, in ellipticals, which have an intense initial starburst, the iron abundance grows very quickly because of type II SNe.

Then type I SNe begin to appear and the $[O/Fe]$ ratio starts to decrease. The same trend of the ellipticals is also present, but milder, in spirals. The opposite behavior can be seen in irregular galaxies that have a lower star formation and present always undersolar values.

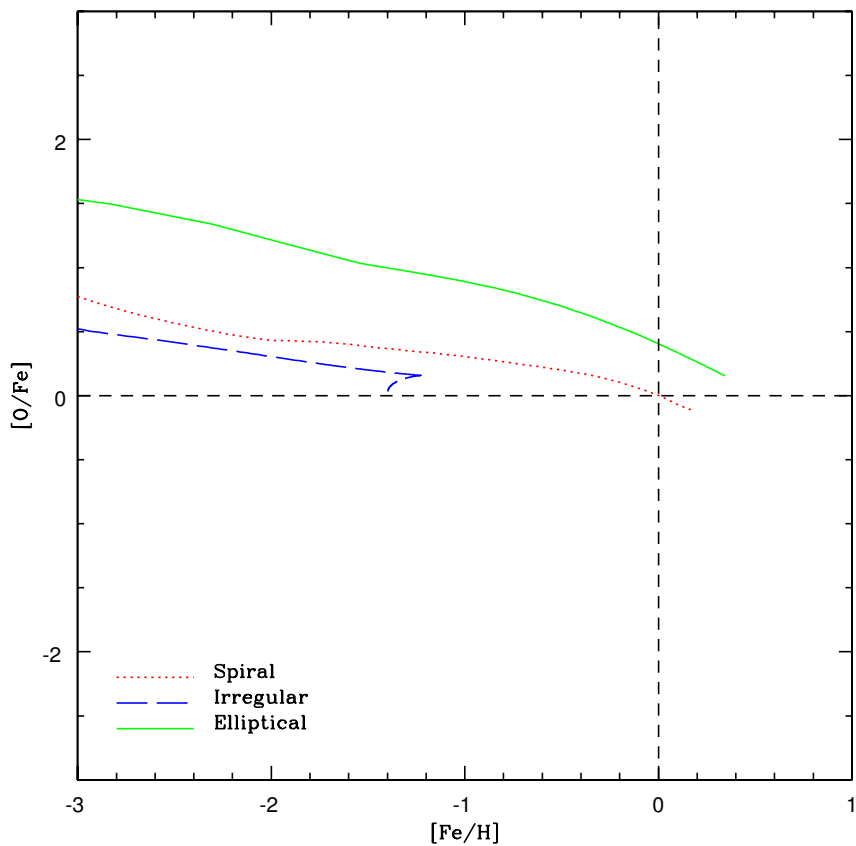

Figure 10: $[O/Fe]$ vs $[Fe/H]$ for a spiral an irregular and an elliptical galaxy.

### 2.4.3 *Metallicity*

An important issue in studying the chemical evolution, is the global metallicity. In the models of chemical evolution, when we speak about metallicity we mean the metallicity of the gas from which stars are born. Metallicity $Z$ is defined as the ratio between the total mass of elements heavier than Helium and the total mass of gas, i.e.:

$$Z = \frac{M_Z}{M_{gas}}. \qquad (2.36)$$

In fig. 11 it is possible to see the redshift evolution of the metallicity of the galaxies used in this work. The galaxies are supposed to form at $z = 10$, having assumed a Lambda-cold dark matter cosmology ($\Lambda CDM$ with $\Omega_0 = 0.3$, $\Omega_\Lambda = 0.7$ and $h = 0.65$).

In the plot it is possible to note that higher metallicities belong to ellipticals with values that are $\sim 4$ times the solar value (equal to 0.0134, from Asplund et al. 2009). For spirals we consider the mean metallicity of the disk and, as we can see, their present time mean metallicity is roughly twice the metallicity of the sun. Finally irregulars present a metallicity that is lower than the solar value throughout the whole cosmic time.

Finally, to stress the goodness of the models used in this work of thesis, we show in figure 12 the evolution of the (solar averaged) metallicity of galaxies of different morphological types compared with metallicities of *quasars (QSO)*, *damped Lyman-alpha (DLA)* and *Lyman-break galaxies (LBG)*. Quasars are thought to be hosted in massive spheroids, while DLA are usually connected to irregulars and LBGs to low mass spheroids. As we can see from figure 12 the predicted evolution of the metallicity of irregulars is in good agreement with data from DLAs which are supposed to be hosted by irregular galaxies or, in the outer part of disks of spirals (Dessauges-Zavadsky et al. 2007), on the other hand also the data from QSO are in agreement with their assumed hosts i.e. high mass ellipticals (Pipino et al. 2011). As we said, LBGs are believed to be hosted by low mass ellipticals and this is the reason why the data are not in agreement with our model for ellipticals (whose baryonic mass is $10^{11}$ $M_\odot$). A less massive model of elliptical would give a better agreement between model and data.

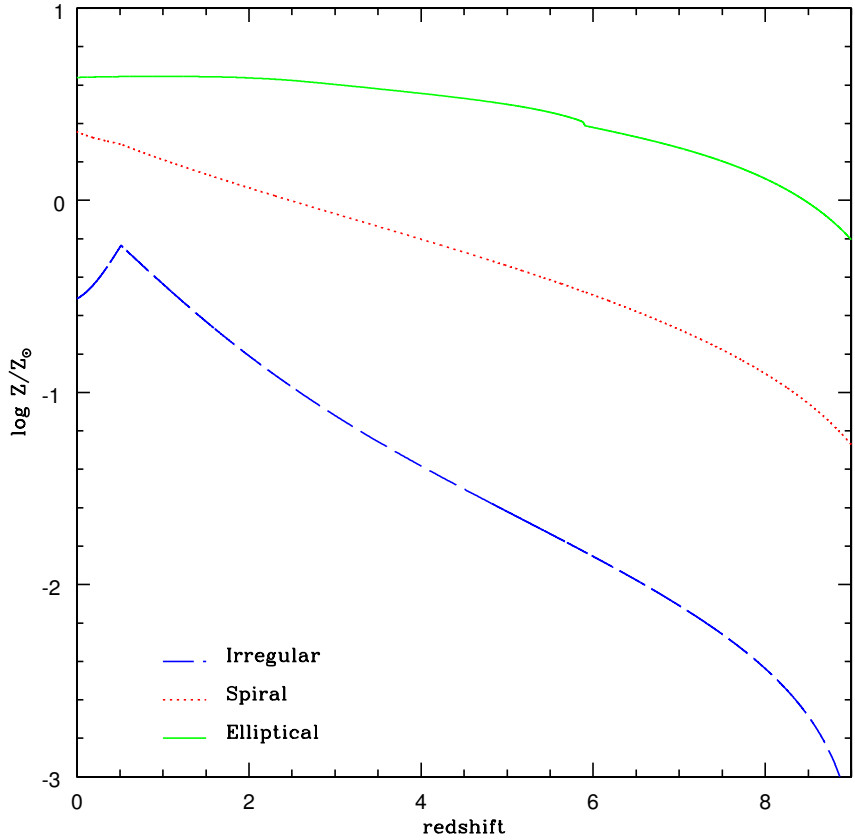

Figure 11: Metallicity vs. redshift for an irregular, a spiral and an elliptical galaxy.

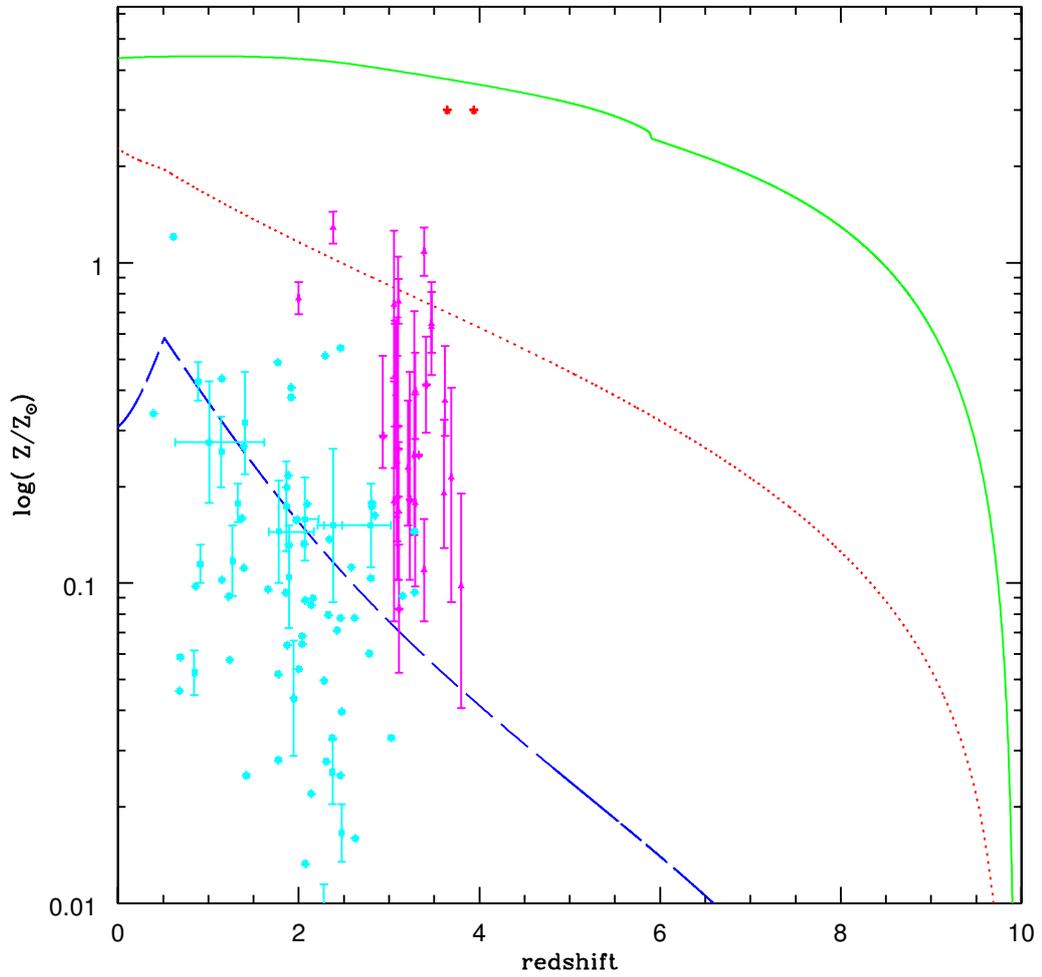

Figure 12: Comparison between the predicted metallicity for galaxies of different morphological types and metallicities of DLA (cyan dots), QSO (red stars) and LBG (magenta triangles). Data are from Pettini et al. 2001, Kulkarni and Fall 2002, Möller et al. 2004, Meiring et al. 2007, Mannucci et al. 2009, Simon and Hamann 2010, Pipino et al. 2011.

# THE PHOTOMETRIC EVOLUTION OF GALAXIES

By means of stellar population synthesis models, it is possible to calculate the evolution of the spectra, magnitudes and colors of galaxies, starting from the evolution of the metallicity, the stellar IMF and the SFR. To study the evolution of the spectrophotometric properties of galaxies, we use the spectro-photometric code GRASIL (Silva et al. 1998). In this chapter we will give a description of the code and point out some results concerning luminosities and spectra of the galaxies. We will also briefly introduce the important question arising from dust extinction and how it can affect measurements.

## 3.1 OVERVIEW

The integrated light coming from galaxies, provides us with fundamental informations about the stellar IMF and the past star formation history, both of which are the basis of any galaxy evolution theory. It is through the collected electromagnetic radiation, that a wealth of properties of galaxies can be deduced. In this context it is of fundamental importance to study properly the evolution of magnitudes, colors, metal content of gas and stars, and other quantities that can provide us a comprehensive study of galactic evolution. To study such properties it is important to estimate the amount of evolution that a galaxy has undergone from a given cosmic epoch to present time. Evolution has to be separated into two components: *intrinsic* that is produced by the birth and aging of the stellar populations, and *apparent* which is produced by the cosmological redshift $z$.

Stellar population synthesis is the mean through which the study of both of these components can be done; in fact it allows us to predict the expected amount of both of these components, under different scenarios. This technique consists in the modeling of the spectral energy distribution *(SED)* emitted by specific stellar populations, and represents the best up to date technique to interpret the light radiated by galaxies. Several groups, developed through the years different evolutionary population synthesis models like Buzzoni (1989), Jimenez et al. (1998), Bruzual and Charlot (2003). All these codes are designed to reproduce the spectral features of galaxies of all types, at any epoch. Although they may vary from code to code, in general, the adjustable parameters of any photometric model are the stellar IMF $\varphi(m)$, the SFR $\psi(t)$ and the metallicity of the gas $Z(t)$. The basic astrophysical ingredients used in these models are *the stellar evolutionary tracks* and *the spectral libraries*, either obtained in an empirical way or based on stellar atmospheric models. The evolutionary track of a star is computed, once the mass of the star and its metallicity are fixed. After this, is then possible, thanks to the stellar evolution theories, to calculate the functions $T_{eff}(M, Z, t)$ and $L(M,Z,t)$ describing the evolution of the effective temperature ($T_{eff}$) and of the luminosity $L$. Effective temperature and luminosity describe parametrically in the Hertzsprung-Russel (fig. 13) diagram the evolutionary track covered by stars of different masses. Generally the stellar evolutionary tracks consist of libraries describing all the stellar evolution phases, from main sequence up to all post main sequence phases.

### 3.1.1 *Main Equations for a Single Stellar Population*

A *Single Stellar Population, (SSP)* is defined as a population of stars born in an instantaneous burst of star formation, with the same initial chemical composition (since they were born at the same time from the same gas) and are characterized

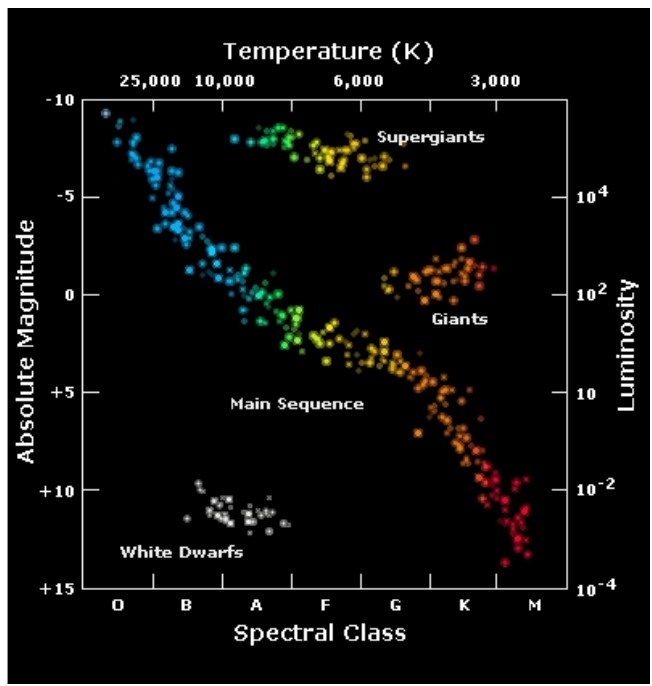

Figure 13: The HR diagram. In the HR diagram the effective temperature of the stars is plotted against absolute magnitude. Stars tends to occupy only specific regions of the diagram depending on their mass and chemical composition. Most of them occupy the diagonal region, the so-called *Main Sequence (MS)*, where stars are burning hydrogen in their cores and where they pass most part of their lives. On the right of the MS from bottom to top we find the Giants branch and the Supergiants branch where we can find stars in a later evolutive stage which are burning heavier elements, than *H*. At the bottom left side of the diagram we find the white dwarfs which present low luminosities and high temperatures and represent the final stage of the evolution of stars ranging from 6 to 8 $M_\odot$. The Sun is located in the middle of the main sequence.

by the following parameters:

- Age;

- chemical composition (i.e. X and Y);

- IMF

A typical example of SSP are globular clusters. Galaxies are not a SSP, because they hold stars of different age and chemical composition so they are a *Composite Stellar Population (CSP)*, i.e. a "sum" of different SSPs. After their formation, it is assumed that all the stars in the SSP evolve passively. Calling $l_\lambda(Z, M, (t_0 - t))$ the luminosity at the wavelength $\lambda$ of a star characterized by a metallicity $Z$ and an IMF $\varphi(M)$, then the luminosity of a SSP at wavelength $\lambda$ can be written as:

$$L_{SSP,\lambda}(Z, t_0 - t) = \int_{M_{min}}^{M_{max}} \varphi(M) l_\lambda(Z, M, t_0 - t) \, dM,$$ (3.1)

where $M_{min}$ and $M_{max}$ are the smallest and the largest stellar mass in the population, namely the lower and upper limits of the IMF. As said before a CSP will consist of a sum of different SSPs, formed at different times, with a luminosity at an age $t_0$ and at a particular wavelength $\lambda$ given by:

$$L_\lambda(t_0) = \int_0^{t_0} \int_{Z_i}^{Z_f} \psi(t - t_0) L_{SSP,\lambda}(Z, t - t_0) \, dZ dt.$$ (3.2)

Where $Z_i$ and $Z_f$ are the initial and final metallicities and $\psi(t - t_0)$ is the SFR at the time $(t - t_0)$, being $t_0$ the age of the SSP.

## 3.2 GRASIL

The GRAphite and SILicate spectrophotometric code GRASIL allows us to perform the photometric evolution of galaxies taking care of the effects of a dusty interstellar medium. Its starting point is a stand-alone chemical evolution code (but, like in our case, also inputs from external codes are allowed) that follows the evolution of star formation rate, metallicity and gas fraction. This is computed thanks to a grid of integrated spectra of simple stellar populations of different ages and metallicities, in which the effects of dust extinction around AGB stars are included. Once these quantities have been deter-mined from the chemical evolution code, the estimate of the SED at the any time $t$ is computed. The building blocks of galaxy models are the library of isochrones. GRASIL is based on the Padova stellar models (Bertelli et al. 1994) with SSPs spanning a range from 1 *Myr* to 18 *Gyr* for what concerns ages, and $Z = 0.004$, $0.008$, $0.02$, $0.05$, $0.1$ for the metallicities (keeping the relative proportion of the metals equal to the solar partition). At this point the spectral synthesis is performed summing up the spectra of each stellar generation provided by the SSP of the appropriate age and metallicity, weighted by the *SFR* at the time of the stars birth, like in eq. 3.2.

### 3.2.1 *The Geometric Distribution*

The peculiarity of GRASIL, is to perform an accurate treatment of dust. The effects of dust are accounted by solving the equations of radiative transfer and in this way it is possible to predict the effect of dust attenuation in the whole spectrum of the galaxy. The effects of the dust depend on the relative distribution of stars and dust. Here the galaxy is supposed to have both azimuthal and planar symmetry relative to the equatorial plane and the dust is divided into three components:

- Star forming molecular clouds *(MC)*;

- free stars (i.e. stars escaped from molecular clouds);

- diffuse gas (cirrus).

The spatial density of stars and MCs (fig. 14) is treated independently but both are supposed to be distributed in the same way. This distribution can be chosen in different ways:

$$\rho = \begin{cases} \rho_0 \cdot e^{(-R/R_d)} \cdot e^{(-|z|/z_d)} & \text{for disklike systems} \\ \rho_0 \cdot [1 + (r/r_c)^2]^{-\gamma} & \text{for spheroidal systems} \end{cases} \qquad (3.3)$$

where $R_d$, $z_d$ and $r_c$ are the scale lengths for the equatorial plane, for the polar axis and the radius of the core respectively. The first two are free parameters.

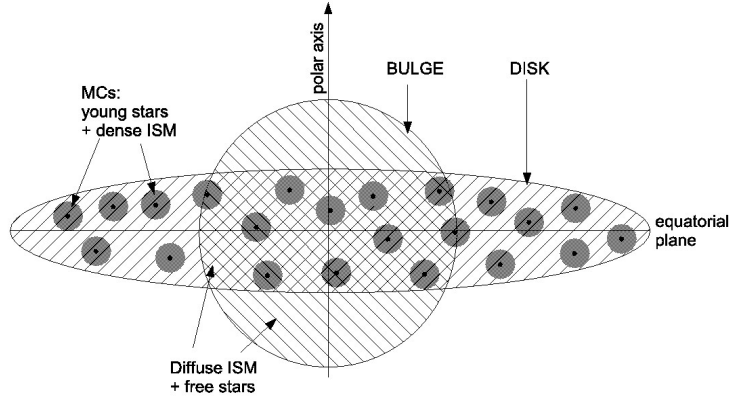

Figure 14: Scheme of the components included in the model and their adopted geometry (from Silva et al. 1998).

Besides the geometry described above, it is possible to set a more general distribution composed of a disk+bulge system. For this work an exponential profile has been chosen for spirals, while ellipticals and irregulars have been modeled through a King profile. The code is build up to receive as an input the fraction of gas in molecular clouds $f_{mc}$ and then this fraction is divided into smaller clouds whose radius ($r_{mc}$) and mass ($m_{mc}$) can be set. Stars are born inside these clouds and then, they escape the cloud (or equivalently the molecular cloud is destroyed) in a time interval $t_{esc}$ that is the timescale of the process, which is inferred to be longer in starburst than in normal disk-like galaxies. Mass and radius of the molecular clouds are very important because, on their basis, the optical depth (namely the fraction of radiation absorbed or scattered along a path through a partially transparent medium) $\tau_{mc} = (m_{mc}/r_{mc}^2)$ is defined, and solving the radiative transfer equations the emerging spectra is obtained.

3.2.2 *The Dust Model*

The luminosity of the stellar population embedded in gaseous regions is affected by obscuration effects by dust grains. In general dust tends to absorb the UV light emitted by young stars and to re-radiate it in the IR-sub mm bands, thus altering the intrinsic spectra of galaxies. This introduces a new problem of degeneracy in the study of an SSP since an observed red stellar population could result from an old age, from an high metallicity or from the presence of high quantities of dust, along the line of sight. To avoid this problem, it is important to take into account the effects of dust obscuration. Several authors proposed different solutions in photometric codes. For example Calzetti (2001) proposed a *"screen"* model in which the absorbed flux of a stellar population, behind a screen of dust is given by:

$$F_a(\lambda) = F_I(\lambda)e^{-\tau(\lambda)}, \tag{3.4}$$

where $F_a(\lambda)$ is the obscured flux, $F_I(\lambda)$ is the intrinsic, unobscured flux at the wavelength $\lambda$ and $\tau(\lambda)$ is the optical depth of the dust screen. In this case $\tau(\lambda)$ depends on several quantities among which the most important is the *extinction curve* that describes how differently the light emitted at various wavelengths is sensitive to dust effects. Different curves exist based either on observations or derived empirically. Another approach is to establish a common extinction (i.e. set an expected level of obscuration) and consequently correct all the data.

The peculiarity of GRASIL is to perform a detailed treatment of the effects of dust obscuration by solving the equations of radiative transfer, once chemical history and geometrical features of the galaxy are given. Derive a standard model of dust, it is not a simple task, since the effects of dust on radiative transfer depend on the physical and chemical properties of the grains as well as the geometrical arrangement of dust-stars, all of them being function of the environment in which grains happen to live. We may distinguish:

- Stellar outflow dust;

- dust in the diffuse ISM;

- dust in molecular clouds and around young stellar objects.

Different absorption is caused by different compositions of the dust. The main features indicate the presence of silicate grains (9.7 and 18 $\mu m$) and of aromatic polyciclic hydrocarbon (PAH) ($3 - 13\ \mu m$). In GRASIL the abundance and the size distribution of grains are calculated starting from the model of Draine and Lee (1984), while their optical properties have been computed by Laor and Draine (1993). In the code the size distribution of graphite and silicate grains is given by:

$$\frac{dn_i}{da} = \begin{cases} A_i n_H a^{\beta_1} & if\ a_b < a_{max} \\ A_i n_H a_b^{(\beta_1 - \beta_2)} a^{\beta_2} & if\ a_{min} < a < a_b \end{cases} \tag{3.5}$$

where $a_{min}$ and $a_{max}$ are the minimum and maximum radius of the size of the grains, $\beta_1$ and $\beta_2$ are the power law indexes from $a_{max}$ to $a_b$ and from $a_b$ to $a_{min}$ respectively, $a_b$ is the break radius between the two power laws, $n_H$ is the number density of $H$ nuclei and $A_i$ are the abundances of graphite or silicates. For PAH, their optical-UV absorption cross-section ($\sigma_{PAH}$) has been calculated using a combination of 6 different $PAH$ mixtures, while in order to compute the IR emission also their heat capacity is taken into account. In figure n. 15 it is possible to see a typical example of a SED obtained with GRASIL in which along with the spectra ranging from radio to UV is possible to see which are the main sources of absorption by dust and which part of the spectra they affect.

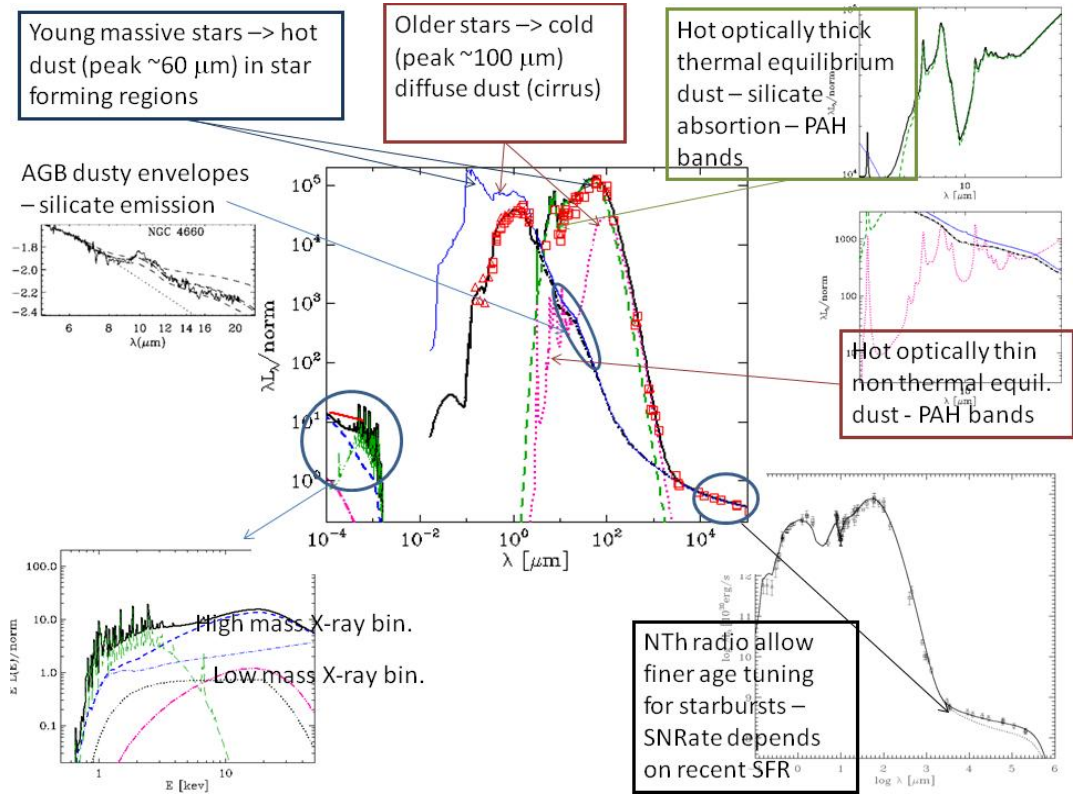

Figure 15: Schematic summary of the sources mostly contributing to the different spectral ranges.

## 3.3 PHOTOMETRIC RESULTS

The GRASIL spectrophotometric code, allows us to calculate the SED of CSP of a fixed age, starting from the output data of a chemical evolution code. This means that GRASIL is able to follow in detail the effect of age-selective extinction (i.e. the observed trend of younger stellar populations which are more affected by dust obscuration) (Fontanot 2005) Given these inputs it is then possible to calculate the corresponding magnitudes in eight different bands (U, B, V, R, I, J, H, K) and the corresponding spectral energy distribution of the galaxy. In what follows we will show the SEDs of the galaxies obtained by using the chemical evolution models described in chapter 2. In particular we adopt the star formation histories described in sec. 2.4.1 together with the results of the chemical evolution models concerning metallicity, Type I and Type II SN rates, gas, stars and infall mass. Before doing that we show in table 4 the main parameters of GRASIL used in this work, for which we have followed the prescriptions of Schurer et al. (2009). A brief description of the meaning of the parameters can be found in the caption.

| Parameter | Spiral | Elliptical | Irregular |
|---|---|---|---|
| $t_{esc}$ [1] | 2 | 250  (0) | 2 |
| $r_c^*$ [2] | ... | 0.21  (0.21) | ... |
| $r_d^*$ [3] | 2.35 | ...  (...) | 1 |
| $z_d^*$ [4] | 0.14 | ...  (...) | 0.5 |
| $r_c^d$ [5] | ... | 0.21  (0.42) | ... |
| $r_d^d$ [6] | 5 | ...  (...) | 1 |
| $z_d^d$ [7] | 0.1 | ...  (...) | 0.5 |
| $M_{mc}$ [8] | $10^6$ | $10^6$  ($10^6$) | $10^6$ |
| $R_{mc}$ [9] | 16 | 16  (16) | 40 |

Table 4: Parameter of GRASIL used in this work. [1] Timescale for the evaporation of MCs, in *Myrs*. [2]Radial scale-length of the stellar component in the bulge, in *Kpc*. [3]Radial scale-length of the stellar component in the disc, in *Kpc*. [4]Vertical scale-length of the dust component in the disc, in *Kpc*. [5]Radial scale-length of the dust component in the bulge, in *Kpc*. [6]Radial scale-length of the dust component in the disc, in *Kpc*. [7]Vertical scale-length of the dust component in the disc, in *Kpc*. [8]Total gas mass in each Mc, in $M_\odot$. [9]Radius of each MC, in *pc*. For the elliptical, the values in bracket refers to the passive phase, while the values outside the bracket refers to the bursting phase.

In figure 16, 17 and 18 it is possible to see the time evolution of the luminosity of the galaxies whose characteristics have been defined in chapter 2, in U (centered at 3650 Å), B (centered at 4450 Å) and K (centered at 21900 Å) band. *B* and *K* band are particularly important since the first one is dominated by hot, massive, young stars while the second one is dominated by low mass, long living stars which contribute to the bulk of the stellar mass. In addition to this, *K* band is also important, since in this band is easier to recognize the effects of the evolution of the metallicity. U band is important also for young stars, since newborn stars emit especially at UV wavelengths but it suffers for strong obscuration by dust.

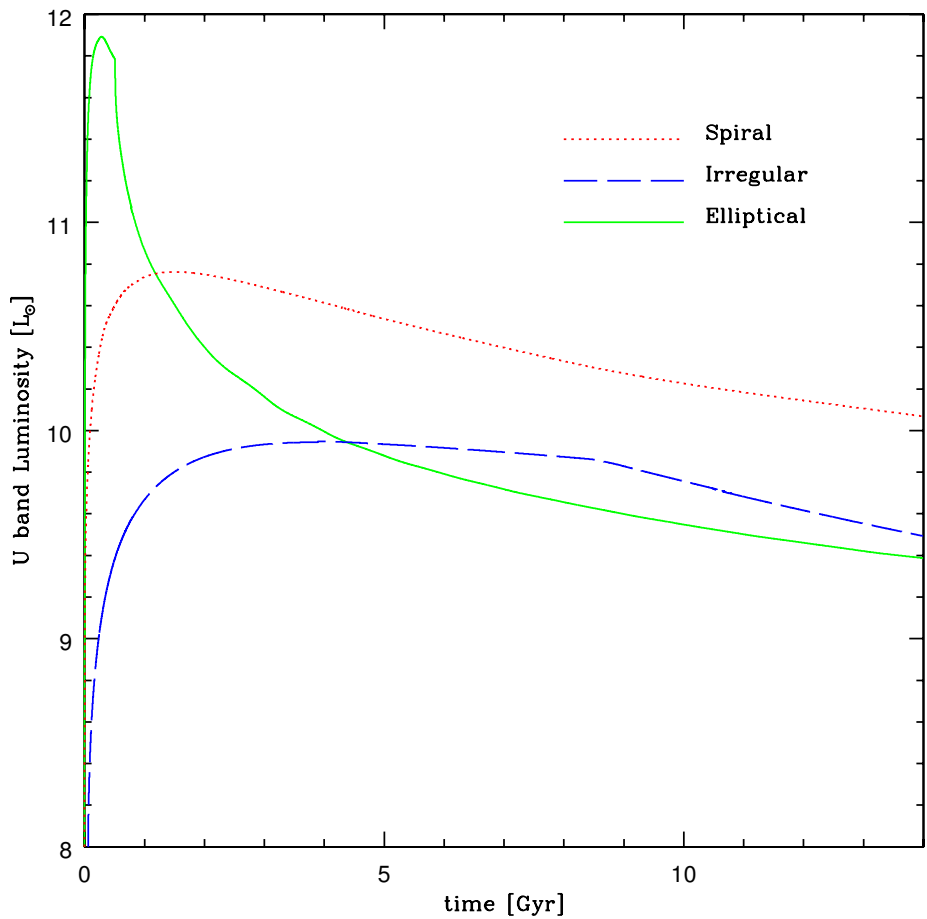

Figure 16: The time evolution of the luminosity in the U band (centered at 3650 Å) for a spiral (red dotted line), an irregular (long dashed blue line) and an elliptical (solid green line) as predicted by GRASIL.

From the same figure it is possible to note that, in each band, elliptical galaxies dominate in the first stages of the evolution, i.e. when they are still forming stars. At later stages (after ∼ 0.6 Gyr) ellipticals stops to form stars and begin their passive evolutionary phase. From now on their luminosity abruptly decreases letting spirals to be the most luminous objects in U and *B* band (due to their ongoing star formation) while in the *K* band this effect is less prominent. With GRASIL it is also possible to study the evolution of the magnitudes in various bands. This has been done in this work and in figure 19 and 20 it is possible to see the evolution of the $B - V$ and $V - K$ colors for the galaxies of different morphological type as predicted by the spectrophotometric code using the results coming from the chemical evolution models we describe before. From these diagrams is possible to evaluate the aging of the stellar population of a galaxy.

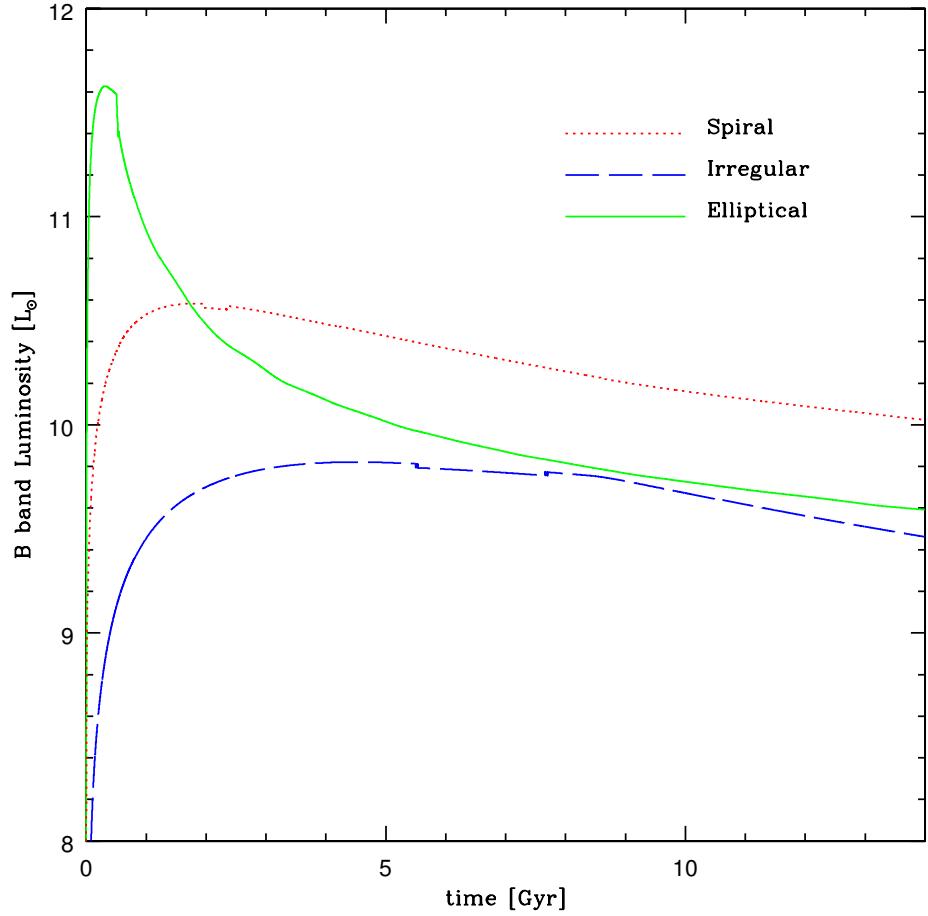

Figure 17: The time evolution of the luminosity in the B band (centered at 4450 Å) for a spiral (red dotted line), an irregular (long dashed blue line) and an elliptical (solid green line) as predicted by GRASIL.

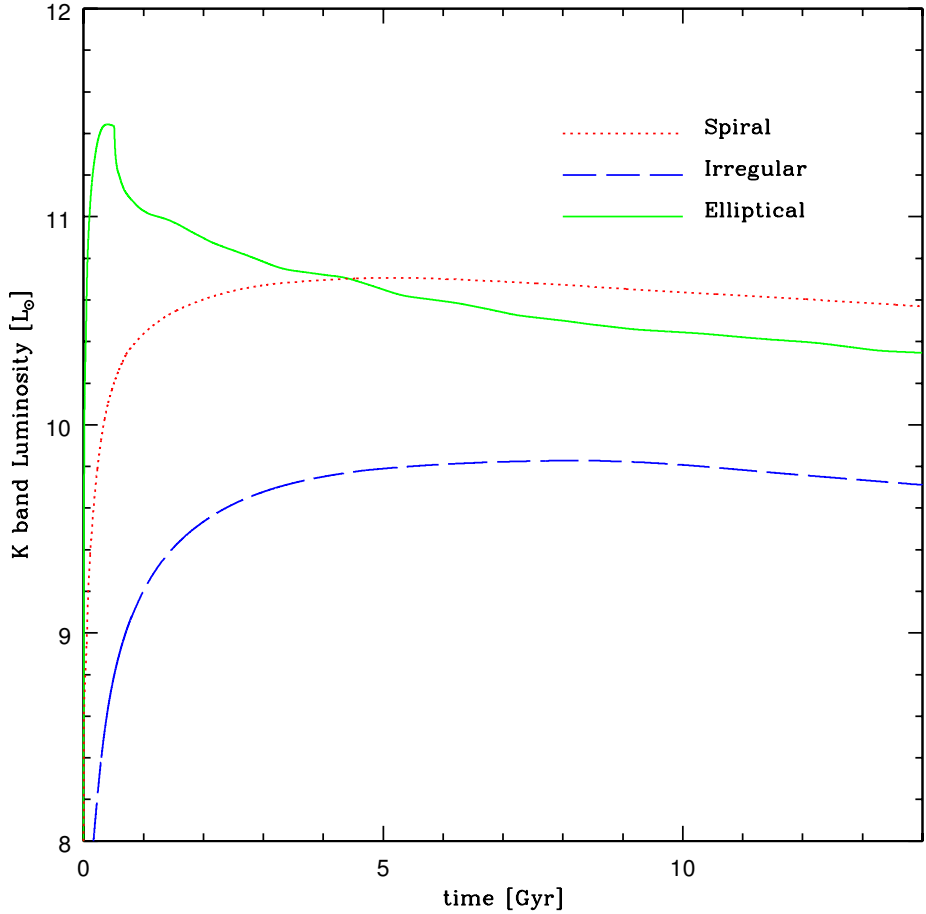

Figure 18: The time evolution of the luminosity in the K band (centered at 21900 Å) for a spiral (red dotted line), an irregular (long dashed blue line) and an elliptical (solid green line) as predicted by GRASIL.

As was said before with GRASIL is possible to compute the spectral energy distribution of a CSP ranging from the UV to the radio. Here we present the SEDs generated with the spectrophotometric code. In figure 21 we show the SED of a spiral galaxy, while in figure 22 and in figure 23 we present the SEDs of an elliptical and of an irregular galaxy respectively.

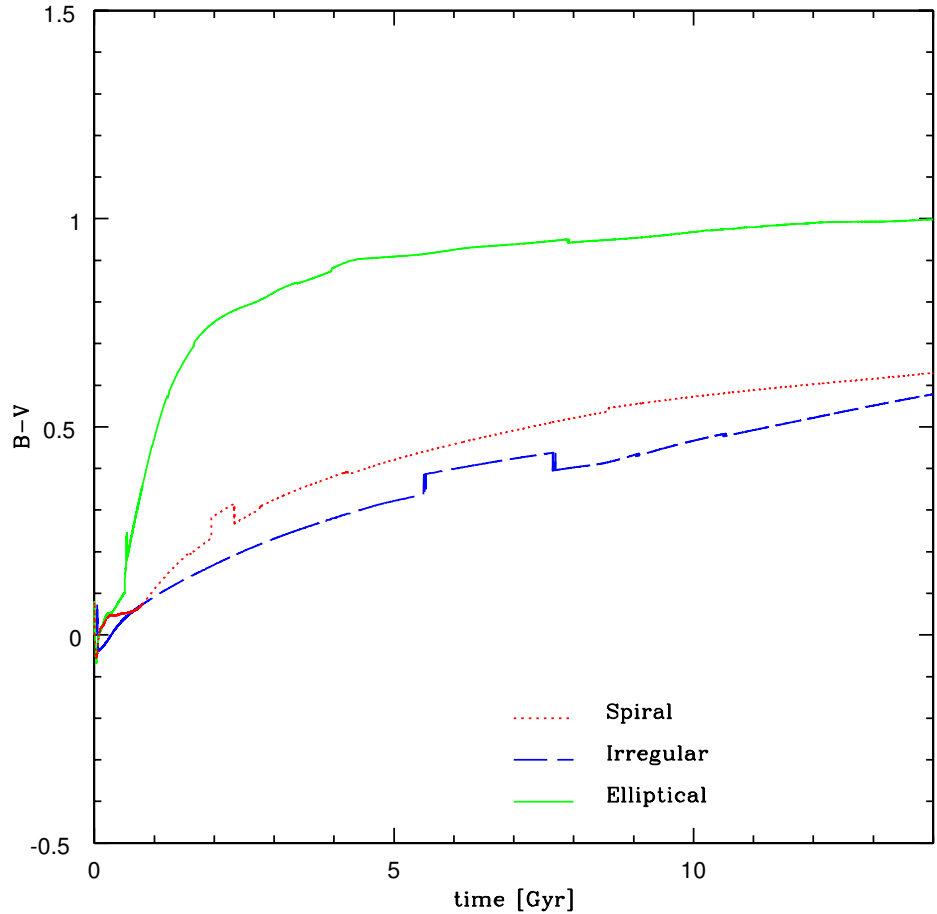

Figure 19: The B-V color diagram for galaxies of different morphological type, as predicted by the spectrophotometric code.

All these SEDs have been generated without taking into account the effect of dust absorption. The reason lies in the fact that all the observational data used in this work are already corrected for dust extinction so, there was no need to produce SEDs with dust absorption features. In this case the result is the pure simple standard stellar population synthesis i.e. the sum of all spectra of stars. It is worth noting that at 0.1 *Gyr* there is a different behavior of ellipticals from spirals and irregulars. In fact, at this epoch ellipticals are experiencing a strong burst of star formation with SFR of about 100 $M_\odot yr^{-1}$ while the other two types of galaxies are producing stars but at a lower level compared to ellipticals. This makes the luminosities of the ellipticals so high at this epoch.

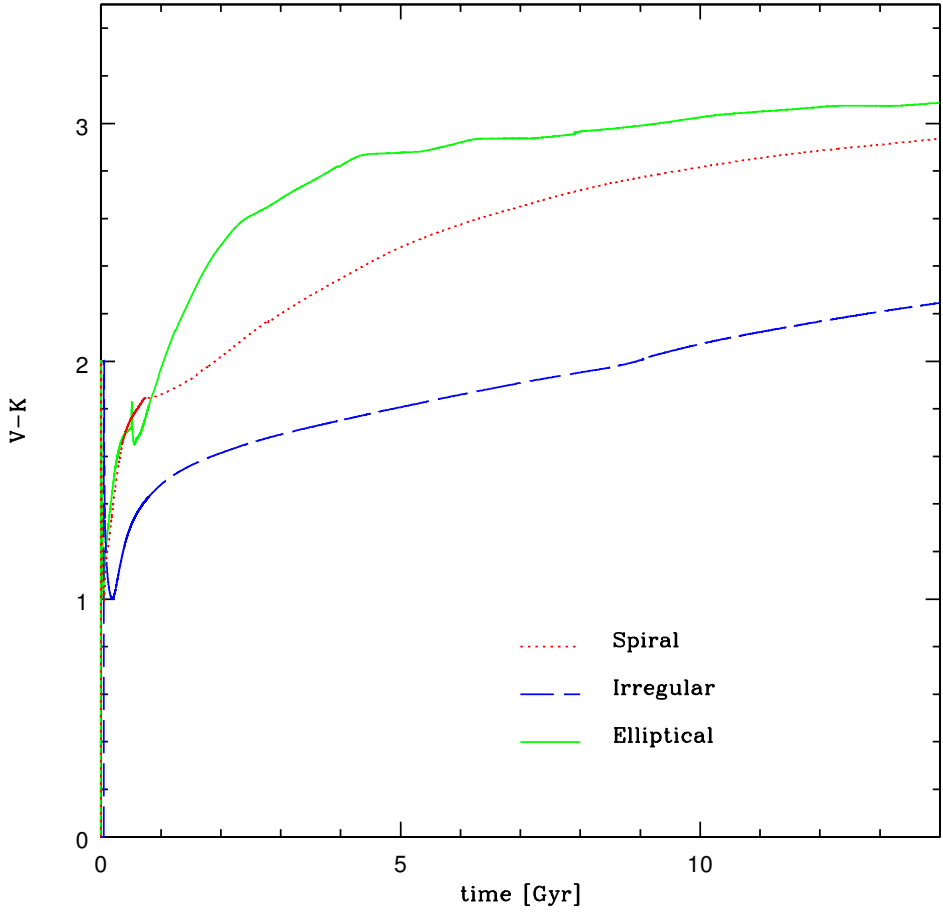

Figure 20: The V-K colors for galaxies of different morphological type as predicted by the spectrophotometric code.

At later times is possible to appreciate the rapid decrease of the luminosity emitted by the ellipticals, while the spectra of spirals and irregulars shows a different behavior due to the different histories of star formation, and suffer a more gentle evolution. Finally to give an example of the dust effects on the SED, in figure 24 we show the differences in the SED of an elliptical galaxy with and without considering dust effects.

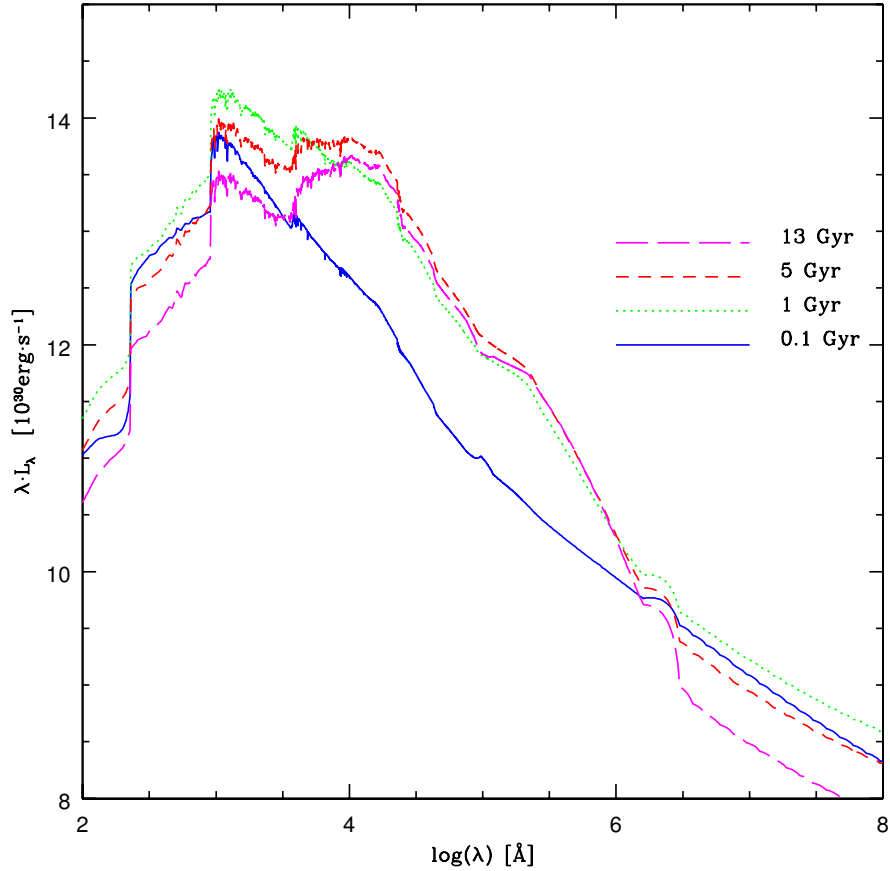

Figure 21: SED of a Milky Way-like spiral galaxy as defined in chapter 2. The SED is plotted at four different epochs: solid blue line 0.1 *Gyr*, green dotted line 1 *Gyr*, red short dashed line 5 *Gyr*, magenta long dashed line 13 *Gyr*.

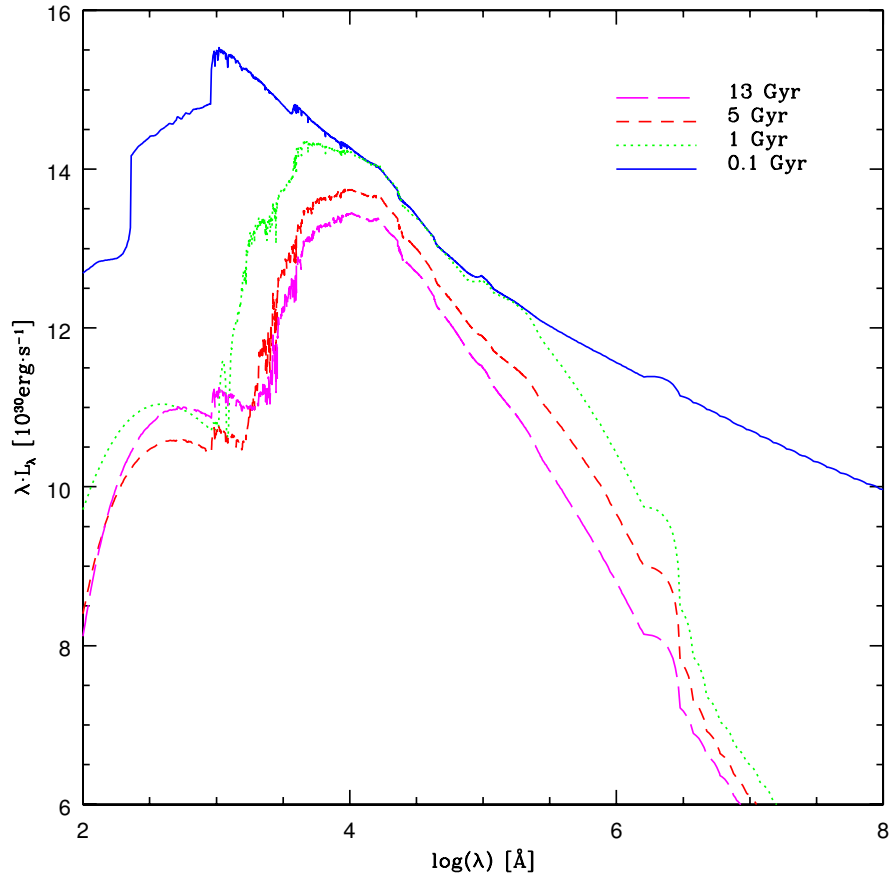

Figure 22: SED of an elliptical galaxy as defined in chapter 2. The SED is plotted at four different epochs: solid blue line 0.1 *Gyr*, green dotted line 1 *Gyr*, red short dashed line 5 *Gyr*, magenta long dashed line 13 *Gyr*.

## 3.4 THE GALAXY LUMINOSITY FUNCTION

To perform the determination of the cosmic star formation rate, it is of fundamental importance the knowledge of the relative distribution of galaxies, of any morphological type, in the Universe. This can be done through the *luminosity function (LF)*, which gives us the distribution of galaxies per unit volume in the luminosity interval $[L, \; L + dL]$. The luminosity function is well reproduced according to the form defined by Schechter (1976):

$$\Phi(L)\frac{dL}{L^*} = \varphi^* \left(\frac{L}{L^*}\right)^{-\alpha} e^{(-L/L^*)}\frac{dL}{L^*}. \tag{3.6}$$

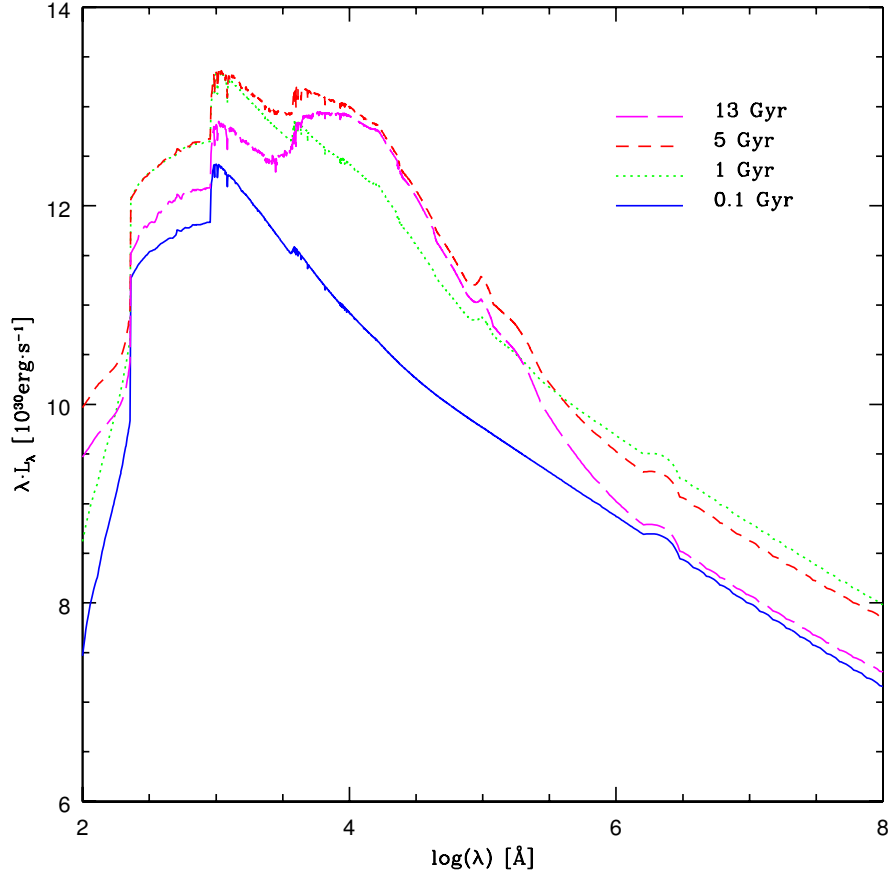

Figure 23: SED of an irregular galaxy as defined in chapter 2. The SED is plotted at four different epochs: solid blue line 0.1 *Gyr*, green dotted line 1 *Gyr*, red short dashed line 5 *Gyr*, magenta long dashed line 13 *Gyr*.

The luminosity function is characterized by three parameters:

- $\varphi^*$ that is the number of galaxies per unit volume;

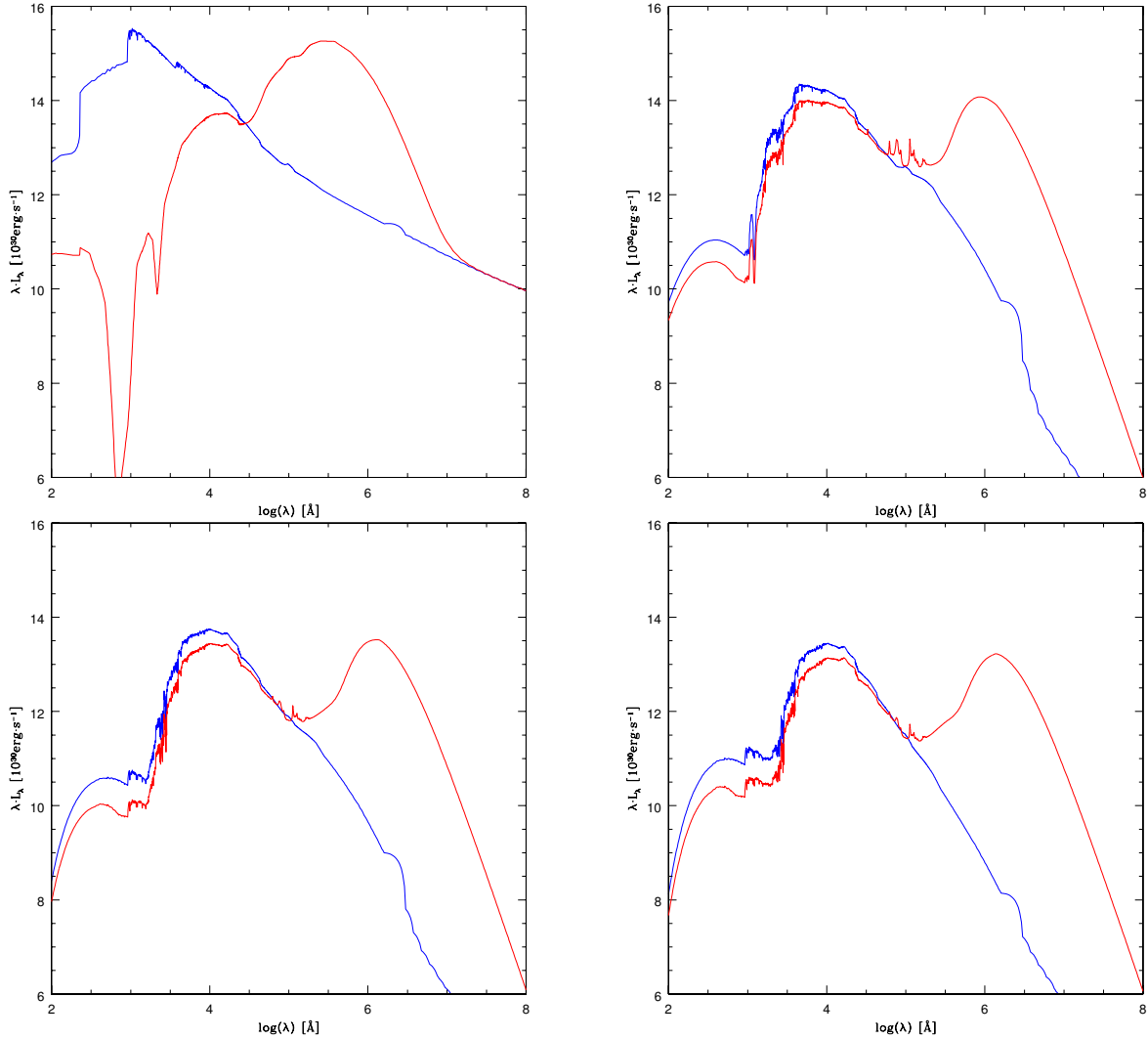

Figure 24: Effects of dust in the SED of an elliptical galaxy at different epochs. Top left 0.1 *Gyr*, top right 1 *Gyr*, bottom left 5 *Gyr*, bottom right 13 *Gyr*. Blue line refers to the model without dust, red line to the model with dust.

- $L^*$ that is the characteristic luminosity that separates bright sources from faint sources, (the break of the luminosity function).

- $\alpha$ that is the slope of the luminosity function.

Different determinations of the local luminosity function exists (see de Lapparent 2003 for a review) and they show several differences, in the sense that the parameters of the Schechter function are not fully constrained by observations. These variations can be partly ascribed to the different selection criteria used in each survey (for example colors criteria, instead of morphological ones). The luminosity function can be measured in several bands, obviously depending on the redshift regime under investigation. At low redshift the local LF can be studied at optical wavelengths, while at high $z$, it is preferable to study it in the UV or IR bands (e.g. Bouwens et al. 2008), since at cosmological distances the optical emission is redshifted towards UV and IR. The luminosity function shows also some changes depending on the sample we are studying. In fact, for example Popesso et al. (2004) on the basis of X-ray data of the *Rosat All Sky Survey (RASS)* (Aschenbach et al. 1981) and of the *Sloan Digital Sky Survey (SDSS)* were able to state that the LF of galaxies in clusters selected in the X-ray band in the ROSAT all Sky Survey is flatter than the LF of field galaxies. In addition to this it can be said that the LF varies depending on the morphological type (see e.g. Baldry et al. (2004) or Marzke et al. 1998) and globally it can be seen as the sum of contributions from the different morphological types.

The parameters of the luminosity function such as $\varphi*$ and $\alpha$ can in principle change as functions of redshift. On the other hand $L^*$ is certainly changing with $z$, since stars evolve in luminosity inside the galaxy at the break.

The consequences of changing the parameters of the LF are of fundamental importance in this work and will be investigated in great detail in the next chapter. In this thesis we make use of the determination of the local luminosity function of Marzke et al. (1998), that has the characteristic to consider separately the different morphological types.

# THE THEORETICAL COSMIC STAR FORMATION RATE

In this chapter, the main results of this work will be presented. Starting from the star formation histories obtained with the chemical evolution codes (chapter 2), and together with luminosities and magnitudes obtained with the spectrophotometric code (chapter 3), we will calculate the cosmic star formation rate, first under the hypothesis of a pure luminosity evolution (PLE) scenario, following the approach of Calura and Matteucci (2003). Then we will test what happens by allowing $\varphi^*$ and $\alpha$ to change with redshift. Throughout this chapter the cosmology adopted will follow the $\Lambda CDM$ paradigm with $\Omega_0 = 0.3$, $\Omega_\Lambda = 0.7$ and $h = 0.65$; where $\Omega_0$ is the matter density parameter, $\Omega_\Lambda$ is the cosmological constant parameter and $h$ is the Hubble parameter.

## 4.1 THE GALAXY LUMINOSITY DENSITY

The first step before we determine the cosmic star formation rate, is the calculation of the *luminosity density (LD)*, which is what is really observed to derive the cosmic star formation rate. The luminosity density, in a given band, is defined as the integrated light radiated per unit volume from the entire galaxy population. It stems from the integral over all luminosities of the observed luminosity function:

$$\rho_L = \int \Phi(L) \left( \frac{L}{L^*} \right) dL. \tag{4.1}$$

By means of the results of the chemical evolution models and of the results of the spectrophotometric code, it is possible to calculate the evolution of the galaxy luminosity density in various bands. At $z = 0$, in the B band, the LDs for the single galaxy types are given by the integral of the luminosity functions of Marzke et al. (1998). At redshift other than zero, for each morphological type we consider the luminosity obtained with the spectrophotometric code.

In figure 25 we can see the predicted evolution of the luminosity density for ellipticals, spirals and irregulars, in the U (centered at 3650 Å), B (centered at 4450 Å), I (centered at 8060 Å) and K (centered at 21900 Å) band.

From this figure it can be clearly seen that, at early times in every band, the luminosity density is dominated by the light emitted by elliptical galaxies. This is obviously due to the fact that they suffer a strong initial burst of star formation which lasts for $\sim 0.6$ $Gyr$. At this time a galactic wind develops and they begin their passive evolutionary phase, so their luminosity in the U and B bands where young, newborn stars emit, begins rapidly to decrease. On the other hand spirals form stars continuously and the luminosity decrease in the U and B band has to be ascribed mostly to the consumption of their reservoir of gas. Finally, also irregulars like spirals, form stars for the entire cosmic time but their luminosity density is lower than the other morphological types due to their low star formation. Their luminosity density is also decreased due to the onset of the galactic wind at later times (after $\sim 10$ $Gyr$).

The I and K bands are dominated by the light emitted by less massive, long living stars. Also in this case, ellipticals are the main sources at early times, while spirals become dominant after $\sim 2$ $Gyr$. Here the difference in the values reached by ellipticals and spirals is less pronounced than in U and B bands. The luminosity density of spirals, after an initial phase of constant increase lasting $\sim 3$ $Gyr$ is observed to decrease slowly in I band and to be quite constant in K band. Irregulars, also in this case have values lower than ellipticals and spirals, that tend to decrease very slowly through the whole cosmic time.

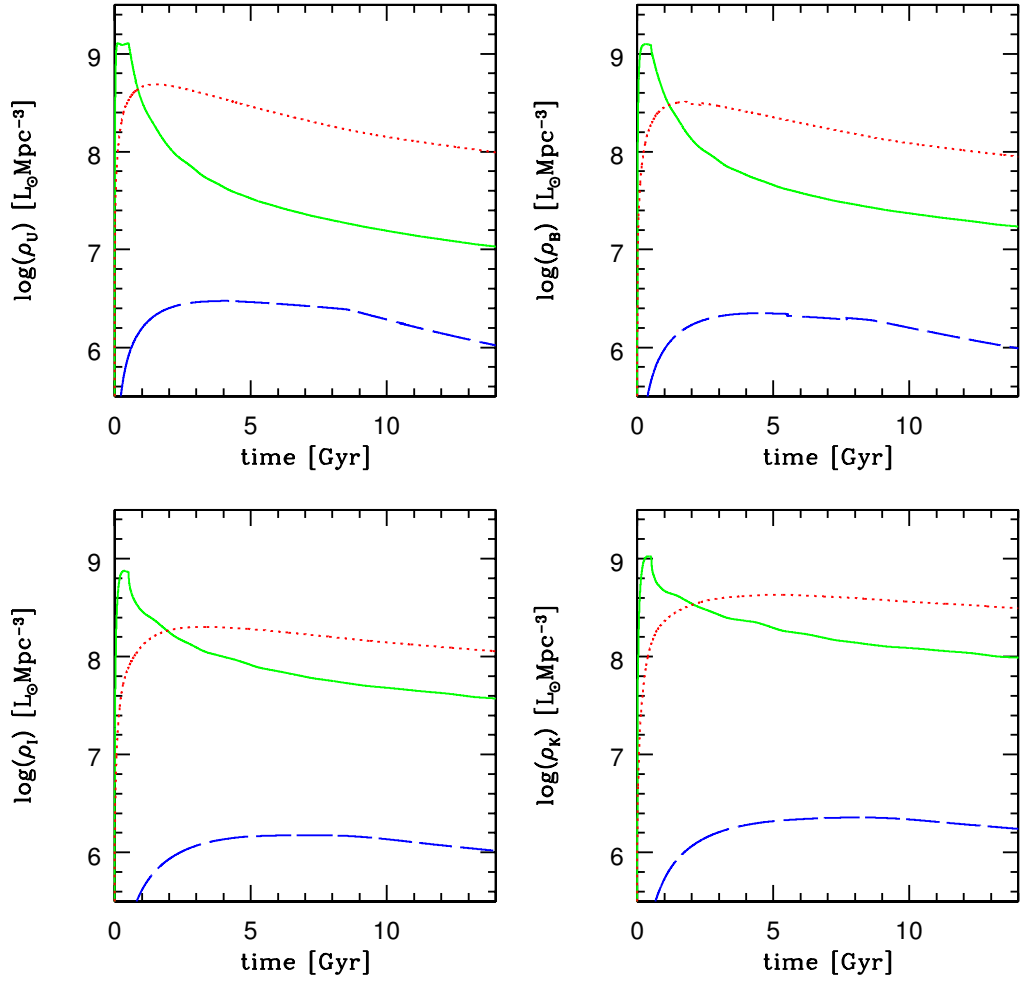

Figure 25: Prediction of the luminosity density evolution for galaxies of different types in U (top left), B (top right), I (bottom left) and K (bottom right) band respectively. Solid green line refers to ellipticals, red dotted line to spirals and long dashed blue line to irregulars.

## 4.2 THE COSMIC STAR FORMATION RATE

The comoving space density of the global star formation rate is known as *cosmic star formation rate (CSFR)*, and it has been measured up to high redshift by means of the comoving luminosity density in various wavelength bands. The physical meaning of the CSFR is of cumulative SFR owing to galaxies of different morphological type, present in a unitary comoving volume of the Universe, convolved with the galaxy luminosity function. At high redshift, we cannot distinguish galaxy morphology but only trace the luminosity density of galaxies. The first measure of the comoving luminosity density of the Universe was done in three wavebands (2800 Å, 4400 Å and 1 $\mu m$) over the redshift range $0 < z < 1$ by Lilly et al. (1996). They found that the comoving luminosity density is increasing markedly with redshift for all the studied wavebands. As a consequence of this, also the CSFR is increasing markedly with redshift up to $z = 1$.

In the following years many studies on the derivation of the cosmic star formation rate have appeared, and at the present data have been collected up to very high redshift, although the data for $z >> 1$ are still uncertain and one of the most important uncertainties is represented by dust extinction at very high redshift. In order to derive the cosmic star formation rate from luminosity densities one has to assume star formation calibrations. The tracers used to derive the CSFR have been the $H_\alpha$, $H_\beta$, $[OII]$, $UV$ and $FIR$ luminosities.

Many studies of the CSFR appeared in the last years and general indicate that the CSFR should decrease for redshift $z > 2$. Generally the behavior of CSFR is interpreted as due to the star formation in galaxies, but as we have already said this is not entirely true since the CSFR may trace also the evolution (if any) of the galaxy luminosity function. However recently, the CSFR has been measured also from Gamma-Ray Burst (GRB) counts and extended up to $z \sim 8$, which corresponds to the estimated redshift of GRB 090423 and it shows no significant decrease of the CSFR. In figure 26 we show the cosmic star formation history determination of Kistler et al. (2009).

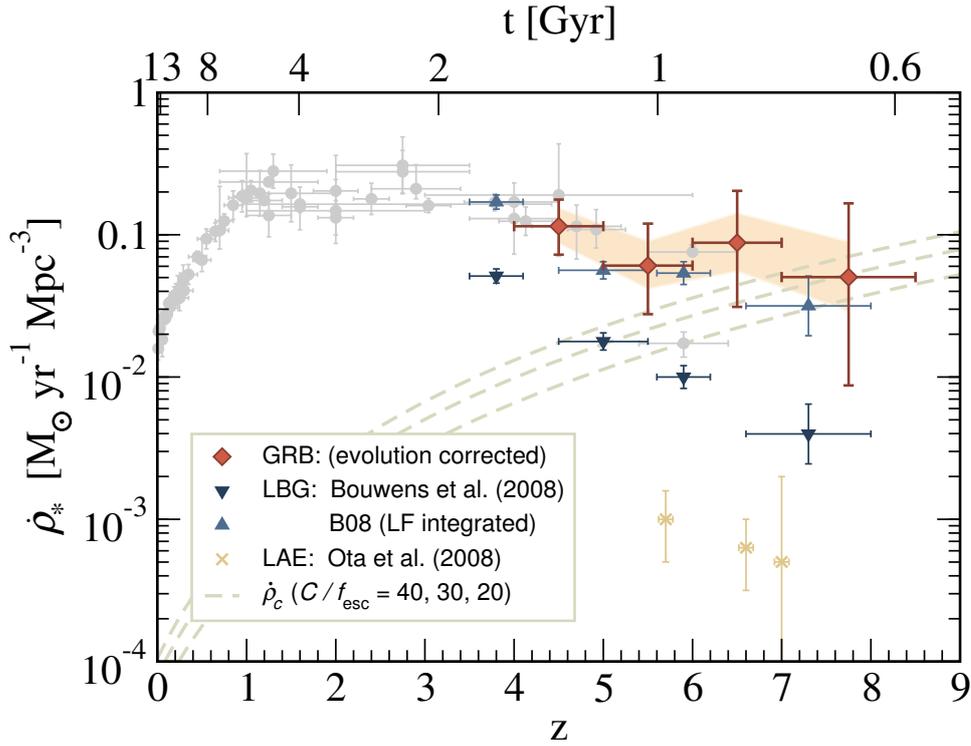

Figure 26: The cosmic star formation history from Kistler et al. (2009). Shown are the data compiled in Hopkins & Beacom (2006) (light circles) and contributions from Lyα Emitters (LAE) Ota et al.(2008). Lyman-break galaxy data is shown for two UV LF integrations: down to $0.2\,L^*_{z=3}$ (down triangles; as given in Bouwens et al. (2008)) and complete (up triangles). Swift gamma-ray burst inferred rates are diamonds, with the shaded band showing the range of values resulting from varying the evolutionary parameter between $\alpha = 0.6 - 1.8$. Also shown is the critical $\dot{\rho}_*$ from Madau et al.(1999) for $\mathcal{C}/f_{esc} = 40, 30, 20$ (dashed lines, top to bottom, $f_{esc}$ is the fraction of photons escaping their galaxy and $\mathcal{C}$ is the clumpiness of the ISM

### 4.2.1 *Observational data of CSFR*

Our models have been compared with the most recent data available in literature. In our work we use the data compilation of Hopkins (2004), Li (2008) and Kistler et al. (2009). In Hopkins (2004) they consider the CSFR determined by several authors using different indicators ($H_\alpha$, $H_\beta$, $[OII]$, 1500 Å, 1600 Å, 1700 Å, 2000 Å, 2800 Å, 15 $\mu m$, 850 $\mu m$ and radio). The data have been converted to a Salpeter (1955) IMF and corrected for dust extinction using a common obscuration correction and an extinction curve. These data show a trend of the CSFR constantly increasing from $z = 0$ to $z \sim 2.5$ where there is a peak. From $z \sim 2.5$ to $z \sim 4$ the CSFR shows a decrease and then a quite constant behavior till $z \sim 6$. The data from Li (2008) are instead based on the detection made by Bouwens et al. (2008) and applying them a dust obscuration factor varying with redshift. From their data a decreasing trend of the CSFR from $z \sim 4$ to $z \sim 10$ is clearly visible. The data from Kistler et al. (2009) that we have used are instead based on the *Swift* GRB sample. It is worth mentioning that these data suggest a more flat trend for the CSFR and also values higher than the one of Li (2008). We also stress that GRBs are found to favour metal poor, sub-L$^*$ galaxies (Stanek et al. 2006, Fruchter et al. 2006). This means that thanks to GRBs we can probably look at galaxies that cannot be seen by any present survey because under the detection limit. This could be the evidence that we are missing a part of star formation and probably underestimating the high redshift CSFR.

### 4.2.2 *The Pure Luminosity Evolution Scenario*

In this section we will describe the determination of the cosmic star formation rate in the framework of a *pure luminosity evolution (PLE)* scenario. With pure luminosity evolution, we mean that galaxies of all morphological types are supposed to form at high redshift and then to evolve only in luminosity and not in number. We also assume that the slope of the LF is constant with redshift. In this context galaxies are assumed to be isolated i.e. the effects of mergers or interactions are irrelevant at all redshifts, and it is assumed that they start forming stars all at the same time.

In this approach, the differences between the various morphological types should be ascribed to their different star formation histories. Early-type galaxies (i.e. ellipticals and *S*0) are dominated by old stellar populations so they must have formed their stars with high efficiency and in a short period, while late-type galaxies (i.e. spirals and irregulars) present, in their colors, evidences of recent star formation sign of a milder variation of the SFR through the cosmic time.

The PLE belongs to the category of the so-called *"backward evolution"* models (Tinsley and Danly 1980); these models start from a well known and well constrained description of the local Universe to reconstruct the evolution of galaxies back to their formation. While they are able to reproduce quite well number counts and observations in various bands at low redshift (He and Zhang 1998, Hill and Shanks 2011), they seem not to be able to reconcile the number of sources at high redshift, with the local number counts (Kitzbichler & White, 2005). On the other hand, the other way to approach the problem is through semi-analytical models, to which we can refer as *"ab initio"* models. Here the evolution of galaxies is modeled starting from collapse and merging of dark matter halos and include gas cooling and heating, star formation and feedback (see for example De Lucia et al. 2006). It must be said that hierarchical models have to face some problems in reproducing adequately at the same time the main observational characteristics of spheroids.

Pure luminosity evolution models require three main ingredients:

1. A parametrization of the star formation history for each morphological type;

2. a cosmological model;

3. the present day luminosity function per morphological type.

The luminosity density has been obtained by integrating the luminosity function over all the luminosities (or equivalently all the magnitudes), and it is described by eq. 4.1

For this work we make use of the B-band luminosity density, because it is a good star formation tracer since hot massive stars emit in this band. At $z = 0$, the luminosity densities for the single galaxy types are simply given by the integral of the observed luminosity function while, at redshifts other than zero we take into account the luminosities calculated by means of the spectro-photometric code. Then, we calculate the B band luminosity function at redshift $z$ according to:

$$\Phi(L_B, z) = \Phi(L_B(z)). \tag{4.2}$$

In this case, as stated before, no merger or interaction between galaxies are assumed, so the parameters of the Schechter function are considered constant, with redshift. In this work we use the parameters set in Marzke et al. (1998), which are based on the *"Second Southern Sky Redshift Survey" (SSRS2)* using data from 5404 galaxies at $z \leq 0.05$. In table 5 we summarize number densities used and slopes of the LF.

The cosmic star formation rate has been computed according to Calura and Matteucci (2003):

$$\dot{\rho}_*(z) = \sum_i \rho_{Bi}(z) \left(\frac{M}{L}\right)_{Bi}(z) \psi_i(z), \tag{4.3}$$

| Morphological Type | $\alpha$ | $\varphi_*$ |
|:---:|:---:|:---:|
| E/S0 | $-1.00^{+0.09}_{-0.09}$ | $4.4\pm0.8$ |
| Spirals | $-1.11^{+0.07}_{-0.06}$ | $8.0\pm1.4$ |
| Irr/Pec | $-1.81^{+0.24}_{-0.24}$ | $0.2\pm0.08$ |

Table 5: Parameters of the Schechter function, from Marzke et al. (1998). First column refers to the morphological type; second column refers to the slope of the luminosity function and the third column indicates the number density of bright objects in a unitary volume of Universe (expressed in units of $10^{-3}Mpc^{-3}$).

where $\rho_{Bi}$ is the B-Band luminosity density, $\left(\frac{M}{L}\right)_{Bi}$ is the B-band mass-to-light ratio and $\psi_i$ (expressed in $yr^{-1}$) is the star formation rate of the $i$-th morphological type.

In our model, all the galaxies are supposed to form[2] at $z \sim 10$. This choice is motivated by the fact that more and more objects are revealed at $z > 5$ (Mobasher et al. 2005, Vanzella et al. 2008), with evidences of a $z \sim 10$ detection Bouwens et al. 2011). The results are shown in figure 27; for comparison we plot in the same figure also the CSFR as computed in Calura and Matteucci (2003).

As it can be seen, both models show some discrepancies when compared with the observational data. In our case the discrepancy begins at $z \sim 0.5$ and is particularly evident in the redshift interval running from $\sim 0.5$ to $\sim 5$. Here we under predict the observed values by a factor of $\sim 3$. This is quite the whole redshift interval in which only spirals and irregulars contribute to the cosmic star formation history. Also the Calura & Matteucci CSFR suffers of the same problem when compared with the most recent data. On the other side, at high redshift, our model seems to overpredict the data. The high observed peak is to ascribe to ellipticals which, in order to reproduce observational features, (like the increase of the $[Mg/Fe]$ ratio and the velocity dispersion $\sigma$) have to form stars in a short time interval (lasting less than 1 $Gyr$, Matteucci 1994). This implies that the SFR of these objects must lie in the range $\sim 100 - 1000 \ M_\odot yr^{-1}$. At $z \sim 6$ ellipticals stop to form stars and the CSFR decreases abruptly. The discrepancies with the model of Calura and Matteucci (2003) can be explained through the differences in the parameters adopted in chemical evolution models and in the different photometric code used.

---

2 We remember that for galaxy formation redshift we mean the redshift at which they begin to form stars.

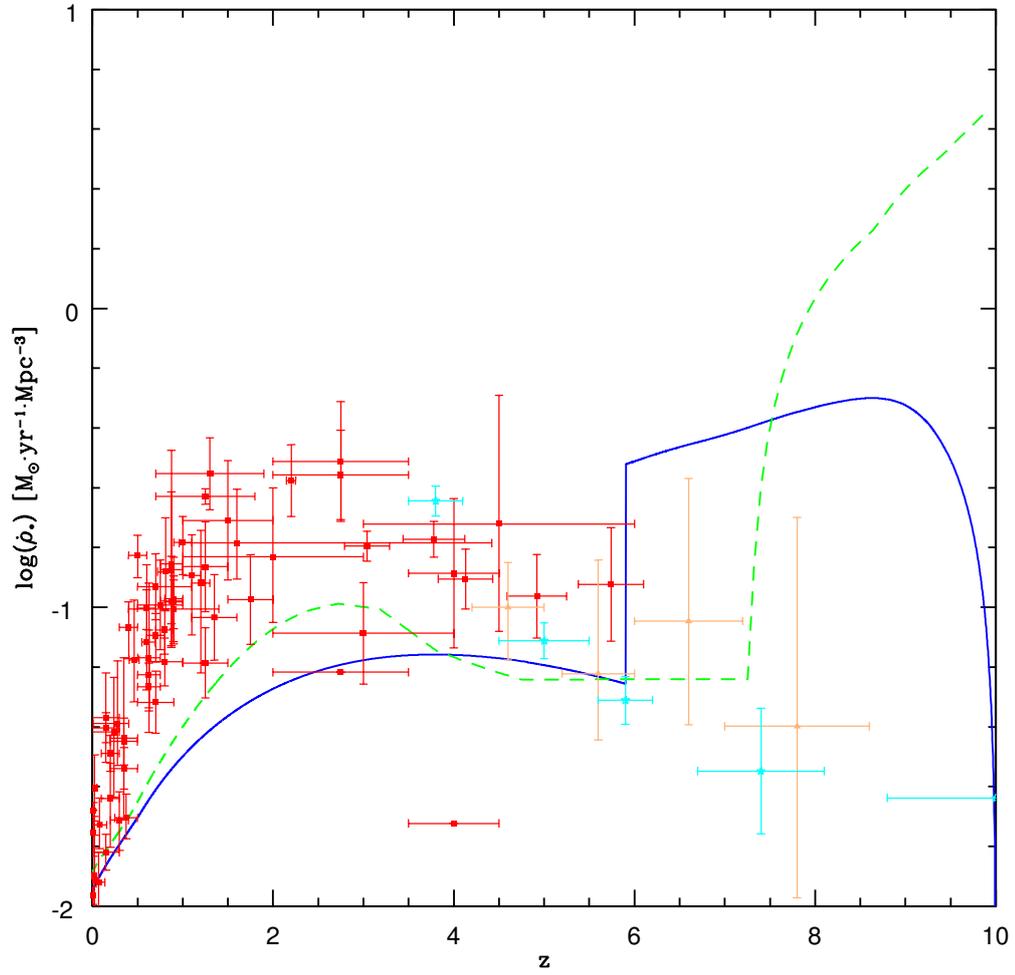

Figure 27: Cosmic star formation history in the case of pure luminosity evolution. Blue solid line, model described in this work; green dashed line model from Calura and Matteucci (2003). Points are taken from: Hopkins (2004) red squares, Li (2008) cyan stars and Kistler et al. (2009) orange triangles.

When we consider the CSFR computed from the observations, we must pay particular attention on how obscuration is treated. In fact, different bands are affected at different level by dust, and is not always straightforward to correctly take into account its effects. A common procedure is to make a dust correction assuming an average level of expected obscuration and uniformly correct all objects. In this case usually, the amount of extinction is determined on the basis of observed average obscuration for similarly selected samples found in the literature. Another approach is to consider extinction curves like for example Cardelli et al. (1989), in ($H\alpha$, $H\beta$ and $OII$) or the galactic obscuration curve of Seaton (1979) and, Calzetti et al. (2000) in the $UV$ (from 1500 to 2800 Å). It is important to stress that different curves give rise to different extinction so, their choice has to be carefully considered. The problem of dust obscuration is even worse at higher redshifts, since determination of absorption begins to be highly uncertain since $z \sim 3$, and still unknown at $z \sim 6$ (Li 2008). Some authors claim that dust extinction at $z \sim 6$ is lower than at $z \sim 3$ (Bouwens et al. 2006) and even absent from $z \sim 7$ Bouwens et al. (2010). To the contrary, Pipino et al. (2011), predict that in ellipticals, at early stages, there is a strong increase in the amount of dust and this can lead us to miss the most of SF in these objects.

The high level of uncertainty means that the peak we obtained at high red-shift could not in principle be an overestimation, but instead there could be an underestimation of the true value of CSFR due to the incorrect treatment of dust. This point of view is also enforced by the cosmic star formation rate determined from GRB. Recently (Ishida et al. 2011), using existing data from *Swift* plus a new GRB recorded at $z = 9.4$ inferred a cosmic star formation rate value at this redshift, higher than the one at present time. This means a factor from 3 to 5 higher than previous estimates based on high redshift galaxies, and so indicating that the decline of the CSFR at higher redshift could not be completely correct, or that it does not mean necessarily a decline in the SFR.

### 4.2.3 *The Number Density Evolution Scenario*

In this section we describe a different approach adopted with respect to the PLE scenario. In what follows we introduce a scenario in which we let the number density of galaxies to evolve with redshift and we will show the results obtained with this choice. It is worth noting that our approach is inserted in the so-called monolithic scenario. In this scenario ellipticals and bulges formed at high redshift as the result of a violent burst of star formation, following a monolithic collapse of a gas cloud. After such a collapse, stellar populations age passively until present time. For what concerns disk galaxies, the monolithic collapse scenario foresees that they formed after bulges, through progressive infall of primordial gas.

The main consequence of relaxing the hypothesis of the constancy of the number density of galaxies is that we allow galaxy interactions. The interactions considered, in order to preserve chemical observational properties, are *dry-mergers*. These are pure dissipationless mergers of stellar systems without any gas and star formation. It has been claimed that giant ellipticals could form, as the result of the occurrence of 1 to 3 major dry-mergers during galactic lifetime (Pipino and Matteucci 2008 and reference therein) and this view is in agreement with the observations. Moreover it has been observed that the fraction of dry-mergers increases with redshift from $z \sim 1.2$ to $z = 0$ (Lin et al. 2008).

For this case we modify the expression of the luminosity function as defined in sec. 3.4, letting the parameter $\varphi^*$ (i.e. the number density of galaxies) to vary with the redshift. The new formulation of the luminosity function is now:

$$\Phi(L)\frac{dL}{L^*} = \varphi_0 \cdot (1+z)^{\beta}\left(\frac{L}{L^*}\right)^{-\alpha}e^{(-L/L^*)}\frac{dL}{L^*}, \tag{4.4}$$

where $\varphi_0 = \varphi^*(z=0)$, i.e. the present time number density of galaxies as defined in sec. 4.2.2. This model is designed in order to reproduce the number densities observed at $z = 0$. We let the parameter $\beta$ to run from $-6$ to $+6$ in steps of 0.2. The choice of this interval has been made in order to explore a set of values in which the results obtained would not "explode" giving us meaningless results and, at the same time, let us to test cases of extreme number density variation. For each value of $\beta$ we calculate the luminosity density using the modified version of the Schechter function. Then, using the star formation histories depicted in chapter 2, we determine the star formation rate density, using the equation 4.3.

In figure 28 we show as an example four cases of the results obtained evolving the number density of galaxies. The cases plotted refer to values of $\beta$ that are equal to: $-4.2$, $-2.2$, $0.8$ and $2.0$. We think that these four cases are quite explicative of the variety of results that can be obtained with the modification explained before. From the plot it is also obvious that it is not possible to reproduce the global cosmic star formation history considering the same behavior for all the morphological types. Doing so would be unphysical since it would be equal to admit that the number density of galaxies of all morphological types, suffers the same rate of variation through the cosmic time in contrast with various studies that we will discuss later.

Always following the number density evolution approach, we decide to impose different evolutions for different morphological types. We perform a care-ful check of the contribution to the total cosmic star formation rate of the different morphological types and on its basis we decide to let only spirals and ellipticals to evolve. Therefore, we consider constant through the cosmic time, the number density of irregulars. The reason stays in their contribution to the total cosmic star formation rate, that is at any redshift quite two orders of magnitude lower than the one of ellipticals and spirals. The chosen values of the $\beta$ parameter for each morphological type are:

- Ellipticals: $\beta = -0.8$;

- spirals: $\beta = 1$;

- irregulars: $\beta = 0$.

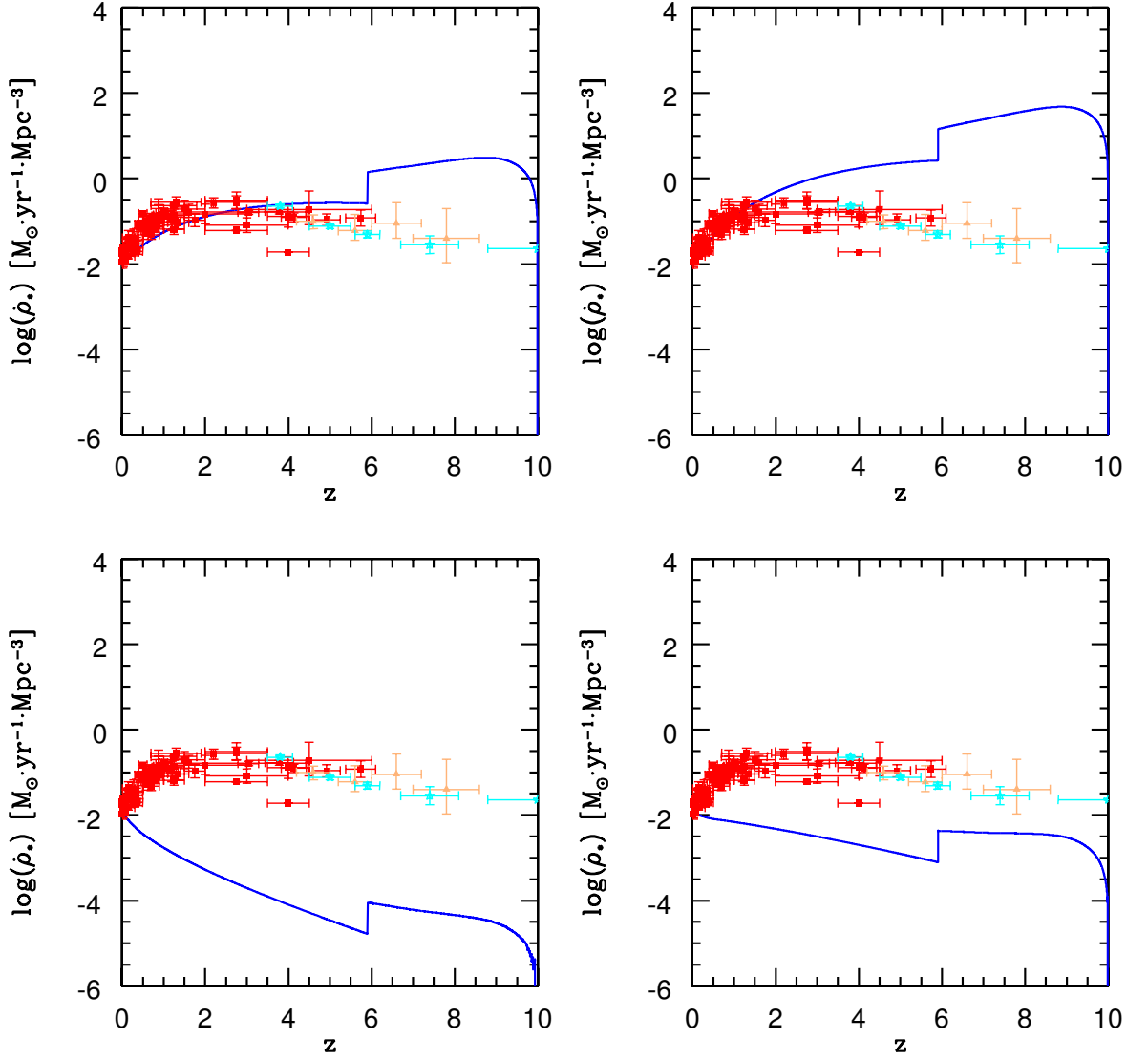

Figure 28: CSFH in the case of number density variation. Here four cases, related to four different values of $\beta$ are plotted, and the values of $\beta$ are the same for all morphological types. Bottom left $\beta = -4.2$, bottom right $\beta = -2.2$, top left $\beta = 0.8$, top right $\beta = 2.0$. In each panel solid line refers to the model results. Data points are the same described in figure 27.

This means that the number density of ellipticals is supposed to increase from $z = 10$ to $z = 0$. The opposite behavior is predicted for spirals, which are supposed to have a linear evolution with redshift. With these assumptions we are considering a picture of the evolution of the Universe in which elliptical galaxies can form also thanks to merging of spirals, as proposed by Toomre and Toomre (1972) and Naab and Burkert (2003). Since we are considering only dry-mergers it means that the merging process quenches the star formation of the progenitors. From our model we can see that the number density of spirals decreases of roughly 60% from $z \sim 2$ to $z \sim 0$, and of 50% from $z \sim 1$ to $z = 0$. This is in good agreement with Boissier et al. (2010) who found a decrease of 50% of Milky Way siblings from $z \sim 1$ to $z = 0$, indicating this percentage as a lower limit. In addition to merging, the same author address these phenomena also to the formation of groups of galaxies.

In figure 29 we show our best model in the case of number density evolution. From the plot we can see that the peak observed in the PLE scenario completely disappears. The model seems to better reproduce the high redshift trend if the error bars are taken into account. Also in the redshift interval between 0 and $\sim 5$ the agreement is improved even if the curve is still lower than the data.

It must be said that the question concerning variation of the number density of galaxies of different morphological types is still matter of debate, since there is in general no concordance among various authors. For example Totani and Yoshii (1998) found that, parameterizing the comoving number density of $E/S0$ as proportional to $(1+z)^\gamma$, the favored value for $\gamma$ is $-0.8 \pm 1.7$, in perfect agreement with our value. This analysis is valid in the redshift range $0.3 < z < 0.8$, and the large error is due to the limited redshift range of their sample. In addition to this, de Lapparent et al. (2004) using data from the ESS (ESO-Sculptor Survey) found a number density evolution of late-type galaxies following the same parametrization, with $\gamma = 2 \pm 1$. Other studies, for example Lilly et al. (1998) found at $z < 1$ an evolution of large disk structures (with a bulge-over-total ratio $B/T > 0.5$) following a $(1+z)^\gamma$ relation, but with $\gamma = \pm 0.5$, lower than the one founded in our model.

Even more challenging is to asses the evolution at high redshift, since very deep surveys are required and the correspondence between actual galaxies and their high $z$ counterparts is not firmly established. This causes a lost of information about the morphology of the various objects, giving at the end controversial results.

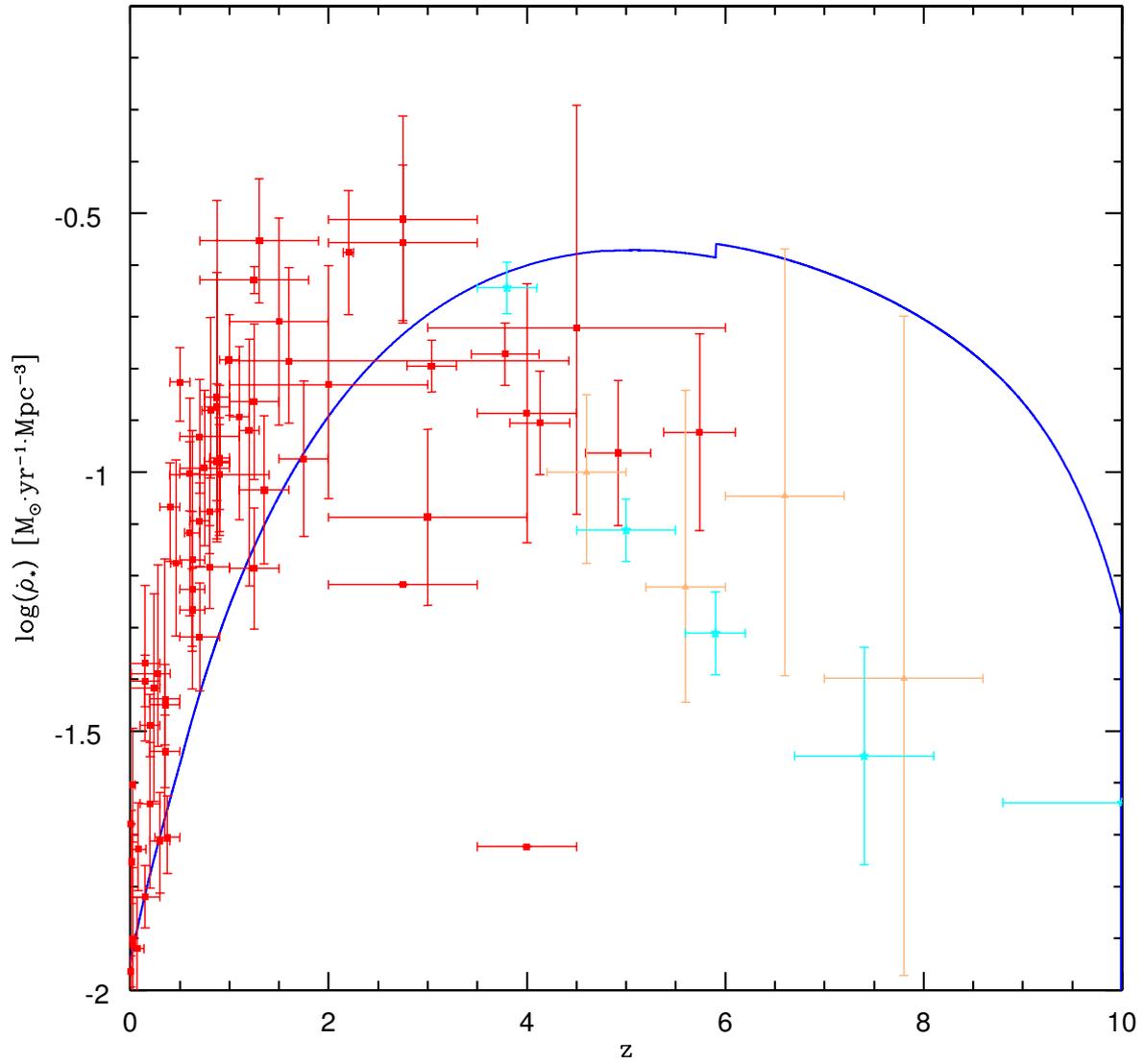

Figure 29: The cosmic star formation history in the Number Density Evolution Scenario. Blue line refers to model results. Data points are the same described in figure 27.

### 4.2.4  *The Variation of α*

Here we introduce a new modification of the Schechter function in which we consider the evolution with redshift of the slope of the luminosity function. We start again from the luminosity function defined in sec. 3.4, and evolving the slope, it assumes the form:

$$\Phi(L)\frac{dL}{L^*} = \varphi^* \cdot \left(\frac{L}{L^*}\right)^{[-\alpha(1+z)^\beta]} e^{(-L/L^*)}\frac{dL}{L^*}. \qquad (4.5)$$

This means that at any redshift we change the slope of the LF thus changing the break that divides the faint end from the bright end. At the end, for each vale of $\beta$ we obtain the convolution of all the LF taken at different redshifts. Like in the previous case, we let $\beta$ to run from $-6$ to $+6$ in steps of 0.2.

From the obtained results we note that this scenario gives us no improvement in the determination of the cosmic star formation history. In fig. 30 we show two examples of the results obtained. It is evident that in this way it is not possible to reconcile our model with data. Therefore, we exclude this possibility of $\alpha$ variation.

### 4.2.5  *Comparison with other models*

In figure 31 we show the results of several determinations of the CSFH that we compare to our model predictions, in particular with the PLE and with our best model with number density evolution. The models we use for this comparison are: Steidel et al. (1999), Porciani and Madau (2001), Cole et al. (2001), Menci et al. (2004) and Strolger et al. (2004). These models are obtained in different ways and we briefly summarize their main characteristics:

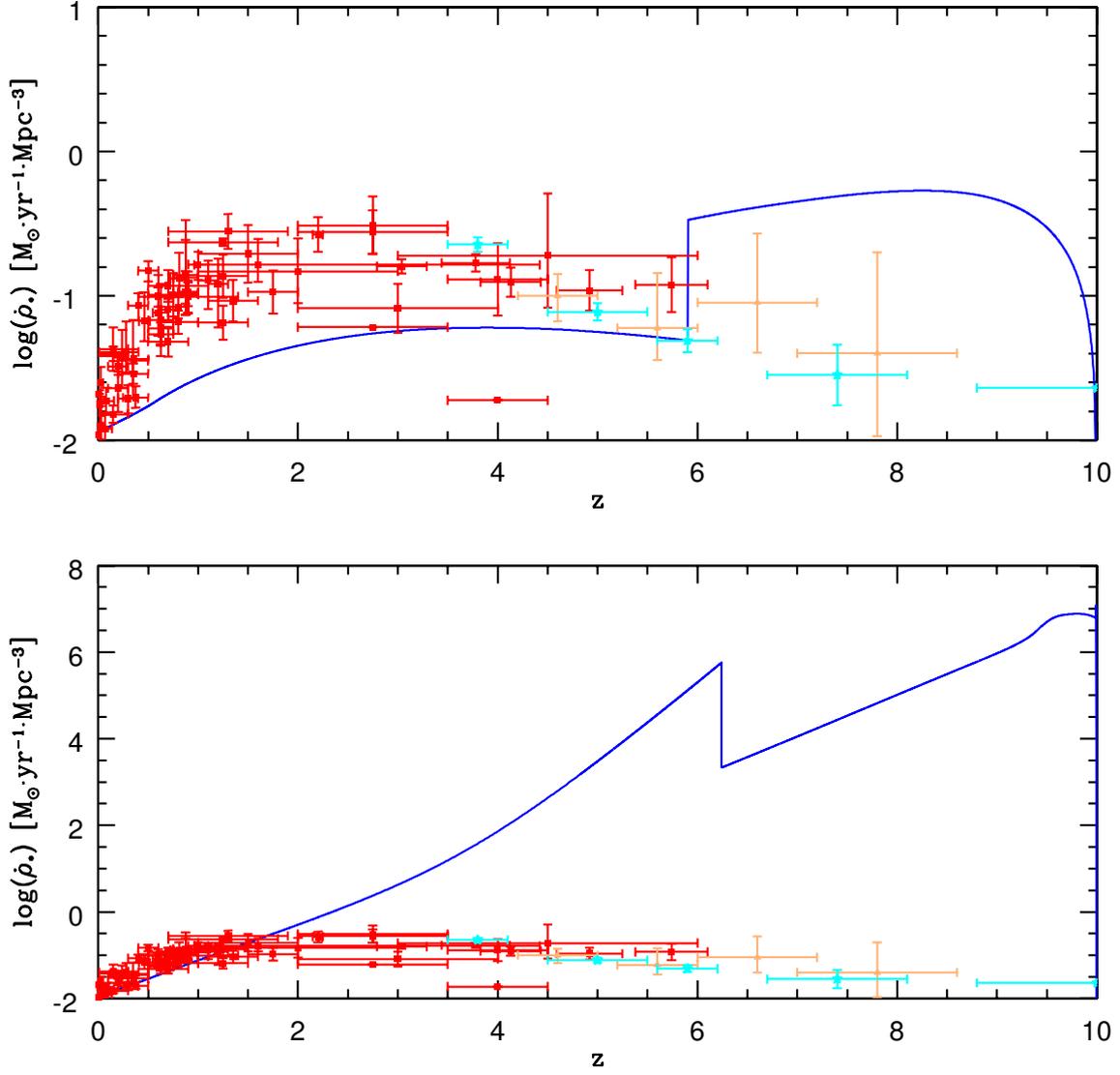

Figure 30: The cosmic star formation history in the case of variation of the LF slope $\alpha$. Here we show the cases corresponding to $\beta = -1.0$ (top panel) and $\beta = 1.0$ bottom panel. Blue lines indicate the results of the model.
Data points are the same described in figure 27. Please note the different y-axis scale adopted for clarity.

- *Steidel et al.*: (magenta dash-dotted line). Their work is based on a sample of *Lyman-break galaxies* founded at $3.8 \leq z \leq 4.5$ and together with data from Madau et al. (1996), Lilly et al. (1996) and Connolly et al. (1997) they infer a CSFH that is quite flat from $z \sim 2$.

- *Porciani & Madau*: (green short-dashed line) Their work is based on the hypothesis that the cosmic rate of gamma-ray burst is proportional to the cosmic star formation rate. They consider different determination of the CSFR from Madau and Pozzetti (2000), Steidel et al. (1999) and Porciani and Madau (2001) and they convert these determination to the $\Lambda$CMD cosmology using the following analytical expression:

$$(\dot{\rho}_*)_{\Lambda CDM} = \dot{\rho}_* h_{65} \frac{[\Omega_0(1+z)^3 + \Omega_K(1+z)^2 + \Omega_\Lambda]^{1/2}}{(1+z)^{3/2}}, \qquad (4.6)$$

where $\Omega_0$ is the matter density parameter, $\Omega_K$ is the curvature parameter (i.e. $1 - \Omega_0 - \Omega_\Lambda$) and $\Omega_\Lambda$ is the cosmological constant parameter. $h_{65} = H_0/65$, where $H_0$ is the Hubble parameter.

- *Cole et al.*: (dark green dotted line) It is a fit of the compilation of Hopkins (2004) data. The fit is expressed by:

$$\dot{\rho}_* = \frac{(a_1 + a_2 z)h}{(1 + (z/a_3)^{a_4})}. \qquad (4.7)$$

- *Menci et al.*: (solid blue line). It is a semi analytical model developed on the basis of the hierarchical clustering. In this scenario dark matter halos can either merge in the central dominant galaxy or retain their identity and became satellite galaxies.

- *Strolger et al.*: (cyan dot dashed line). They consider a modified version of a model of Madau et al. (1998), taking into account the extinction by dust.

$$\dot{\rho_*}(t) = a_1(t^{a_2}e^{-(t/a_3)} + a_4 e^{(t-t_0)/a_3}), \tag{4.8}$$

where $t_0 = 13.47\ Gyr$ is the age of the Universe corresponding to $z = 0$. In the last two models $a_1$, $a_2$, $a_3$ and $a_4$ refers to the parameters of their best fits. From figure 31 it is possible to see that our model with number density evolution shows a complete agreement with the semi analytical model of Menci et al. and, in general, shows a better agreement with the data compared to the models of Steidel et al. and Porciani & Madau. For what concerns the PLE model, this presents lower values than the other models, in the redshift range $[0 - 6]$. From that point instead, it exceeds all the other predictions considered here, as we have discussed already.

*A brief summary*

Summarizing this part of our work, here we try to build up a consistent model to reproduce the cosmic star formation history. We start from models of chemical evolution, make us able to model galaxies of different morphological types. Then using the histories of star formation obtained from these chemical models, we compute the photometric properties of the galaxies. The results of the photometric code have been used to compute the luminosity densities of the various galaxies. Finally, all this information has been put together to obtain the cosmic star formation history. Starting from the pure luminosity evolution scenario, we explore the possibility to evolve the parameter of the Schechter function (i.e. the number density and the slope) introducing a dependence from the redshift. We compute a wide number of models ranging from a strong increase with redshift of the number density and of the slope, to the opposite behavior, i.e. a strong decrease of the parameters mentioned before, from the beginning till now. In the case of PLE we are not able to well reproduce the trend predicted by the data. In fact we underpredict the low $z$ side (roughly from $z = 0$ to $z = 6$) by a factor of 3, while we overpredict the observed trend from $z \sim 6$ to $z$ 10. In the $[0 - 6]$ redshift interval, two are the main causes for this discrepancy: the first could lie in the fact that we choose the Milky-Way as representative of the whole population of spirals. This can lead to underestimate the contribution to the cosmic star formation rate from spirals since we are missing massive disks with higher star formation rates. The second could be the fact that we are not considering starburst galaxies at intermediate-low redshift. Even if there is not a single definition, starburst galaxies can be described as galaxies in which an intense burst of star formation is taking place. The intensity of the burst has to be compared with the mean SFR of the galaxy over its life. Generally, during the starburst phase, the star formation can reach values of thousands of $M_{\odot} yr^{-1}$. Starbursts usually affect dwarf and spirals (McQuinn et al. 2009) and are observed both at low and high redshift. At $z \sim 1$ starbursts are found in $\sim 15\%$ of galaxies (O'Connell 2005) with typical duration of the burst of the order of $\sim 200 - 400$ Myr. Taking into account this class of galaxies can in principle increase the CSFR at low redshift, helping us with the discrepancy found.

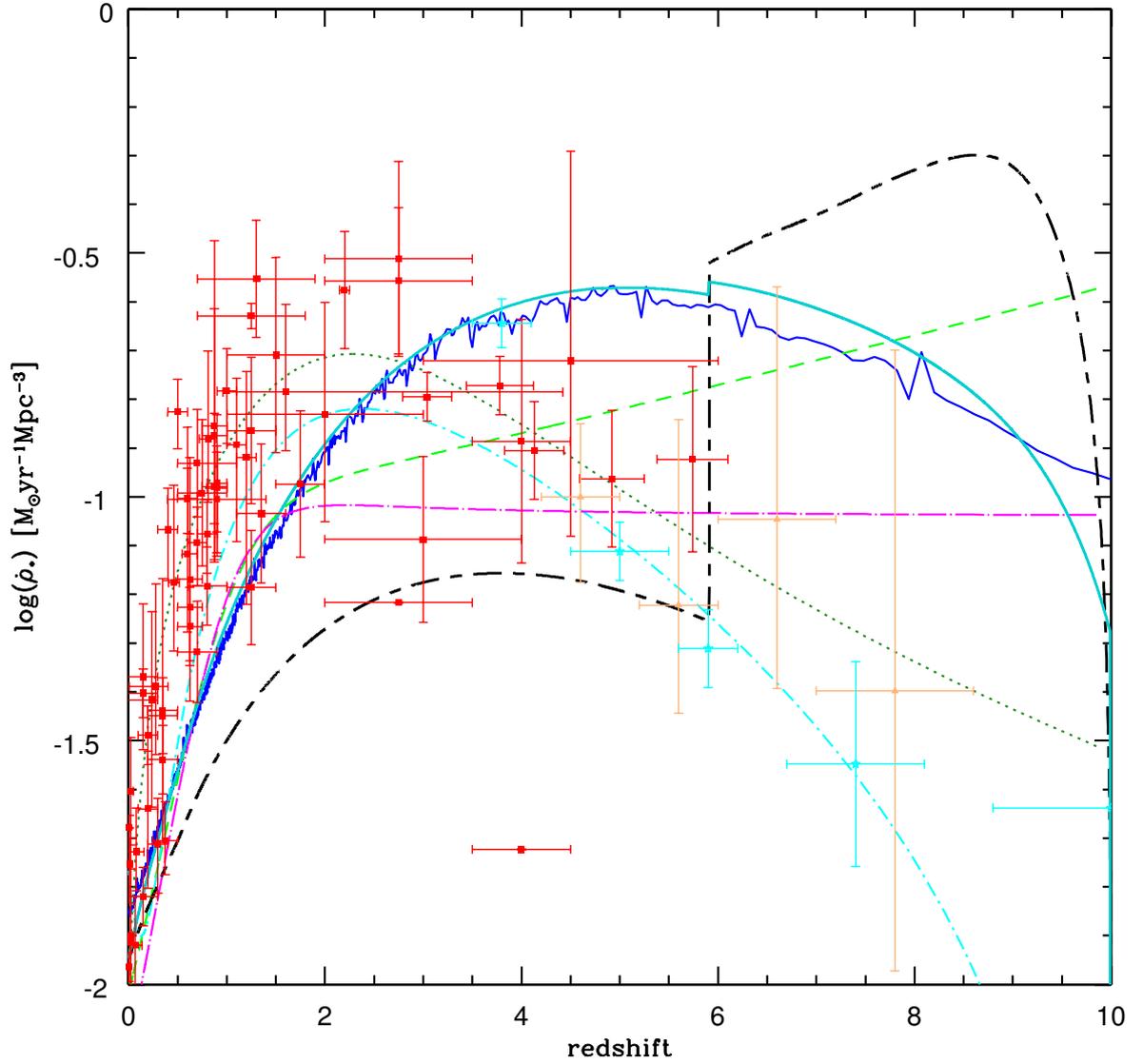

Figure 31: A compilation of determinations of the CSFH, compared to the results of this work. Models are from: Steidel et al. (1999) magenta dash-dotted line, Porciani and Madau (2001) green short-dashed line, Cole et al. (2001) dark green dotted line, Menci et al. (2004) blue solid line, Strolger et al. (2004) cyan dot dashed line, this work (PLE) dark dashed line, this work (number density evolution) turquoise solid line. Data point are the same as fig. 27.

In the $[6 - 10]$ redshift interval, as discussed before, the discrepancy may also arise because of an incorrect treatment of extinction due to dust. In fact, the SF data taken from the literature, usually are obtained converting luminous fluxes or luminous densities into SF through empirical relations. If dust extinction is not taken correctly into account, this can lead to wrong determinations of the SFR and as a consequence of the cosmic star formation rate. In addition to this, the paucity of available data at very high redshift, makes uncertain the determination of the cosmic star formation rate.

These are the discrepancies, and the possible solutions, when discussing the PLE scenario. On the other hand, if we consider the number density evolution scenario the agreement of our best model is better than for the PLE. Although among various authors there is no general agreement upon the rate of evolution of the number density of galaxies, the results obtained with our model better reproduce the low redshift side of the cosmic star formation history, although the CSFR at low redshift is still underestimated. No model of CSFR existing in the literature well reproduces the data at intermediate and low redshift. Also for this scenario the sources of discrepancy discussed for the PLE are still valid. At high redshift we still overestimate the data, but also here the agreement is improved.

## 4.3 THE COSMIC ISM MEAN METALLICITY

Here we compute the evolution of the cosmic luminosity weighted mean interstellar metallicity of the galaxies of different morphological type. With this definition we indicate the mean metallicity of the gas from which stars are born, in a unitary volume of the Universe. The quantity we calculate is the luminosity weighted metallicity. As suggested by Kulkarni and Fall (2002) we compute it through the following expression:

$$\bar{Z} = \frac{\int Z_i(L) L_i \Phi_i(L) \, dL}{\sum_i \int L_i \Phi_i(L) \, dL},$$

(4.9)

where $Z_i(L)$ is the average interstellar metallicity in a galaxy of luminosity $L$, at any given cosmic time, and of the $i$-th morphological type and $\Phi_i(L)$ is the luminosity function of the $i$-th morphological type.

We compute this quantity both in the case of pure luminosity evolution and in the case of number density evolution, using the metallicities obtained with chemical evolution models, together with the local B-band luminosities obtained with the spectrophotometric code. The parameters of the local luminosity functions, are from Marzke et al. (1998), as in the previous paragraphs.

From figure 32 it is clear that in the PLE scenario ellipticals drive the mean ISM metallicity throughout the whole cosmic time. This is due to their higher average metallicity compared to spirals and irregulars. This is an observational fact and in our models the high metallicity of ellipticals is reproduced by means of a strong initial burst of star formation, as we have already discussed in the previous chapters. The spiral galaxies are the second most important contributors to the cosmic chemical enrichment, whereas irregulars give a negligible contribution. We can see that our model is in agreement with Calura and Matteucci (2004) who found that for $z \sim 2$ the main sources of metals production are spirals. It is worth noting that, the decrease of the luminosity weighted metallicity of ellipticals is due to the fact that at $z \sim 6$ in our model ellipticals stop forming stars and from that moment their B-band luminosity abruptly decreases. Also spirals decrease their B-band luminosity but mildly since their star formation lasts for the whole cosmic time. We note that in this case two distinct effects are involved: the metallicity which is increasing with redshift and the luminosity which decreases with $z$.

In figure 33 we show the same as in figure 32 but under the hypothesis of number density evolution. In particular, we show the results of our best model, as described in section 4.2.3. The most evident fact we can note in this case is that now spirals are the main contributors to the total cosmic luminosity weighted mean metallicity. This is in contrast with previous results from other authors, who found that spheroids are the main contributors to the cosmic metallicity. However, we need further investigation to better interpret this result. It has to be considered that this quantity cannot be compared directly with observations, since it is a cosmic average, while measurements are performed in single objects. Another important thing we note is that, in this case the cosmic luminosity weighted mean metallicity of ellipticals is increasing throughout the whole cosmic time. This can be explained by the fact that the decreasing of the B-band luminosity is compensated by the increase of their number density at lower redshift. We also find that our present time value of the cosmic luminosity weighted mean metallicity of the ISM in all galaxies and in each galaxy formation scenario is $\sim 2.4\ Z_\odot$, a value that is slightly higher than $1.3\ Z_\odot$ as found by Calura and Matteucci (2004).

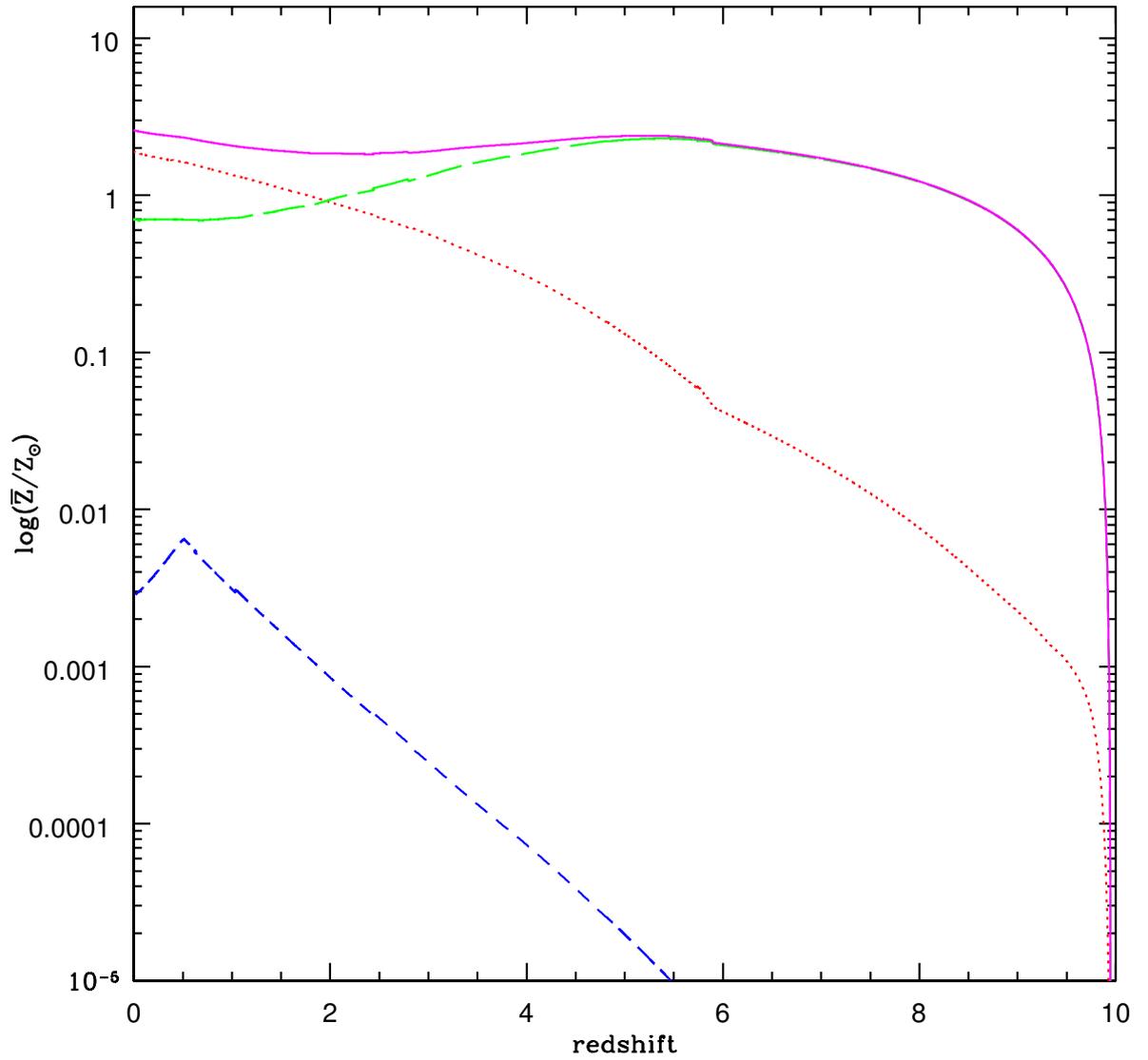

Figure 32: The cosmic luminosity weighted mean metallicity relative to the Sun in the case of pure luminosity evolution. Magenta solid line: total; red dotted line: spirals; green dash-dotted line: ellipticals; blue dashed line: irregulars. All the values are normalized to the solar metallicty $Z_\odot = 0.0134$, from Asplund et al. (2009).

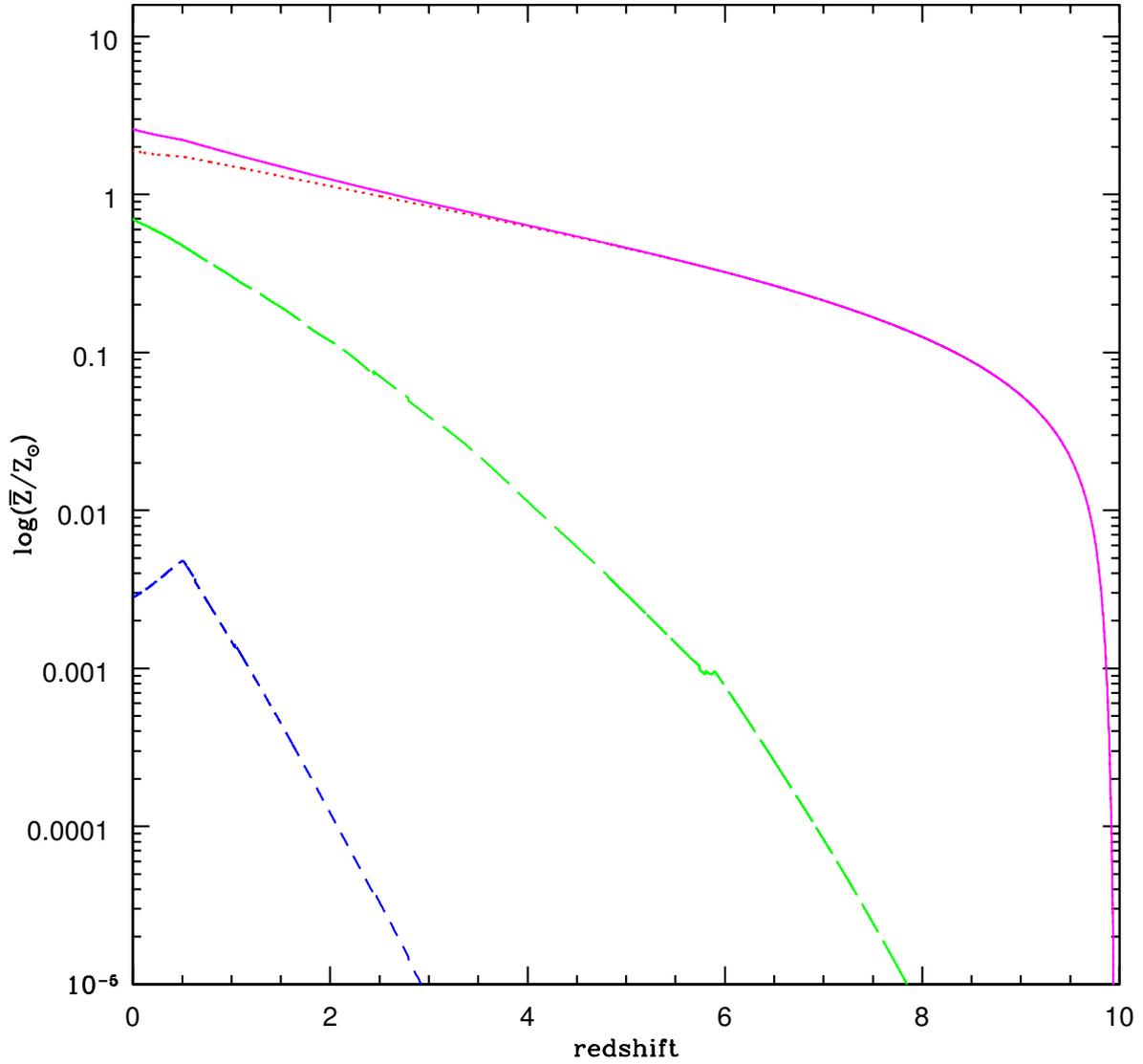

Figure 33: The cosmic luminosity weighted mean metallicity relative to the Sun in the case of number density evolution. Magenta solid line: total; red dotted line: spirals; green dash-dotted line: ellipticals; blue dashed line: irregulars. All the values are normalized to the solar metallicty $Z_\odot = 0.0134$, from Asplund et al. (2009).

This average metallicity is not the global average metallicity of the Universe since we did not consider the metallicity of the intergalactic medium.

CONCLUSIONS

The accurate determination of the cosmic star formation rate is of fundamental importance to establish the correct framework in which galaxies are born and evolve. Its determination is challenging since it requires the knowledge of the star formation history at increasingly high redshifts, and it involves deeper and deeper surveys (like for example *SDDS, GOODS* and the *HDF*). The aim of this thesis is to build up a consistent model to determine the cosmic star formation history of the Universe, through a pure theoretical approach.

In the first part of this work, we develop detailed chemical evolution models for galaxies of different morphological types, namely ellipticals, spirals and irregulars. These models are able to account for the main observational constraints like, among others, present day chemical abundances, star formation rates and SN rates. We consider one galaxy per morphological type (namely, spirals, spheroids and irregulars). Their characteristics are designed to fit the mean properties of the different morphological types. From these models we infer the histories of star formation to be used in the computation of the CSFR.

The histories of star formation then, are included in a spectrophotometric code (GRASIL) to obtain the photometric properties of the modeled galaxies, i.e. luminosities, magnitudes and colors, together with their spectral energy distributions. The chemical and photometric results obtained have been convolved with the luminosity function of galaxies of different morphological types in order to obtain the luminosity densities from which we determine the cosmic star formation rate.

In the second part of this thesis we develop several scenarios to compute the cosmic star formation rate under different assumptions. First, following the approach of Calura and Matteucci (2003) we consider a pure luminosity evolution scenario. In other words, we consider that galaxies evolve only in luminosity, whereas their number densities are assumed constant and equal to the values indicated by Marzke et al. (1998) for the present time luminosity functions. This means that in this scenario we do not consider merging and interaction between galaxies as main mechanisms in galaxy formation. In all our models we assume, for all galaxies, a redshift of formation equal to 10, according to the latest observations of GRB at $z \sim 9.4$ (Ishida et al. 2011) and the detection of a galaxy at $z \sim 10$ (Bouwens et al. 2011).

The results of this model are that at high redshifts ellipticals are the main sources of the cosmic star formation rate, while at lower redshifts ($z < 6$) spirals become the main contributors. The contribution of irregulars is found to be negligible at every time. Comparing our model with a compilation of available data, we obtain that our model underpredicts data at low redshift, while for $z > 6$ we overpredict the observed values.

In this last case the problem could reside in the uncertainties still existing in the CSFR at high redshift and, in particular, in the uncertainties in the dust corrections at high redshift. For what concerns the low CSFR at low redshift a possible explanation could be that we considered a Milky Way like spiral as representative of spirals, and that is an underestimation of the real SFR in the galaxies. Moreover, we do not consider starburst galaxies at intermediate-low redshifts.

Then we consider a scenario in which we evolve the number density of galaxies, i.e. we take into account possible episodes of merging between galaxies. To do this we introduce a dependence from redshift of the number density. We explore a wide range of possibilities for the number density evolution and we find a model which better reproduces the observational data. In particular, this model reproduces the CSFR better than the model assuming a PLE. Our pre-
dicted number density evolution is in agreement with other studies found in
literature (Totani and Yoshii 1998, de Lapparent et al. 2004 and Boissier et al.
2010). We compare our results also with other theoretical models; remarkable
is the agreement of our best model with number density evolution with the
semi-analytical model of Menci et al. (2004).

Finally, we also compute a scenario considering the variation of the slope of
the luminosity function, but this model is rejected because any possible variation
of $\alpha$ does not produce any improvement in the agreement with the data.

In the last part of this work we calculate the cosmic luminosity weighted
mean metallicity of the ISM in all galaxies, both for the PLE and the number
density evolution scenario. In the first case, (PLE) we find that elliptical galaxies
dominate the average metallicity of the Universe at high redshift, and they are
responsible for the production of the bulk of metals, in agreement with previous
work. In the case of number density evolution we find that this is not true since
spirals are the main contributors of metals for the whole cosmic time. This is a
new result and requires further future analysis.

From our work, it is not possible to firmly establish which model has to be
preferred although the one considering number density evolution produces a
better agreement with data. In fact if, from one hand we can say that the PLE
model does not fit well the observational data, on the other hand it produces
chemical results in very good agreement with observations. For what concerns
the number density evolution model, the behavior of the mean metallicity in the
Universe is not so clear, because it attributes the bulk of the metals in the Uni-
verse to spiral galaxies, whose star formation rate is generally mild compared
to that in the spheroids, which host the QSOs where the abundance are much
larger than solar.

Only the availability of more data at high redshift and a better comprehension
of the correct treatment of dust obscuration can shed light on the real trend of the cosmic star formation history and help us to discriminate between competitor models.